\documentclass[a4paper,11pt]{article}
\usepackage{jheppub}

\usepackage[utf8]{inputenc}
\usepackage{graphicx,cancel,float}
\usepackage{amssymb}
\usepackage{textcomp}
\usepackage{amsmath}
\usepackage{tabularx}
\usepackage{bm}
\usepackage{times}
\usepackage{color}
\usepackage{slashed}
\usepackage{multirow}
\usepackage{verbatim}
\usepackage{epsfig,epstopdf}
\usepackage{subfigure}

\allowdisplaybreaks[4]

\def\lsim{\mathrel{\raise.3ex\hbox{$<$\kern-.75em\lower1ex\hbox{$\sim$}}}}
\def\gsim{\mathrel{\raise.3ex\hbox{$>$\kern-.75em\lower1ex\hbox{$\sim$}}}}

\definecolor{orange}{rgb}{1,0.5,0}

\preprint{}
\subheader{\it \today}

\title{Dark magnetic dipole property in fermionic absorption by nucleus and electrons}

\author[a]{Tong Li}
\emailAdd{litong@nankai.edu.cn}
\affiliation[a]{School of Physics, Nankai University, Tianjin 300071, China}
\author[b]{Jiajun Liao}
\emailAdd{liaojiajun@mail.sysu.edu.cn}
\affiliation[b]{School of Physics, Sun Yat-Sen University, Guangzhou 510275, China}
\author[a]{Rui-Jia Zhang}
\emailAdd{zhangruijia@mail.nankai.edu.cn}

\abstract
{The fermionic dark matter (DM) absorption by nucleus or electron
targets provides a distinctive signal to search for sub-GeV DM. We consider a Dirac fermion DM charged under a dark gauge group and with the dark magnetic dipole operator. The DM field mixes with right-handed neutrino and interacts with the ordinary electromagnetic charge current via the kinetic mixing term of gauge fields. As a result, the incoming DM is absorbed and converted into neutrino in final state through the dipole-charge interaction. For the DM absorption by nucleus, the recoil energy spectrum exhibit a peak at $m_\chi^2/2m_N$ for each isotope in the target. XENON1T can probe the DM mass above 27 MeV and the projected constraint on the inelastic DM-nucleon cross section becomes $10^{-49}$ cm$^2$. CRESSTIII with lower energy threshold would be sensitive to the DM mass above 2 MeV. We also check that the contribution from the nuclear magnetic dipole is negligible for $^{131}{\rm Xe}$ target.
The absorption of DM by bound electron target induces ionization signal and is sensitive to sub-MeV DM mass. The involvement of the ionization form factor spreads out the localized recoil energy. We show the future prospect for the constraint on the magnetic dipole coupling from the electron ionization of $^{131}{\rm Xe}$.
}

\arxivnumber{arXiv:2201.11905}

\begin{document}

\maketitle
\setcounter{page}{2}

\newpage

\section{Introduction}
\label{sec:Intro}

The existence of dark matter (DM) has been supported by a number of observation evidences in astronomy. However, the DM particle has not been observed in the terrestrial facilities and its microscopic nature is still unknown. The null evidence of the ordinary DM-nucleus elastic scattering in direct detection experiments encourages us to pay attention to other theoretical hypothesises and search methods.
The inelastic DM is of particular interest among the alternative strategies~\cite{Tucker-Smith:2001myb}. The idea was originally introduced to reconcile the tension between the DAMA annual modulation signature~\cite{DAMA:2000mdu,DAMA:2010gpn} and null result from other direct detection experiments. The mass difference $\delta\sim 100$ keV between the incoming DM and the final state changes the kinematics and thus plays as an important unknown parameter to explain the DAMA signal spectrum. It was pointed out that the inelastic DM transition can be dynamically induced by a dark magnetic dipole under the idea that the DM particle could have an off-diagonal magnetic dipole~\cite{Pospelov:2000bq,Sigurdson:2004zp,Gardner:2008yn,Masso:2009mu}. Moreover, the sizable magnetic dipole of iodine in DAMA's NaI target should be taken into account to accommodate the positive DAMA signal through the dipole-dipole inelastic scattering~\cite{Chang:2010en,Barger:2010gv}.

Apparently, a particular case of inelastic DM is the transition from DM to nearly massless neutrino in which the mass difference is not an additional free parameter any more. Refs.~\cite{Dror:2019onn,Dror:2019dib,Dror:2020czw} recently proposed the idea of fermionic DM absorption by nucleus or electron targets which eventually emit a neutrino in final states, that is
\begin{eqnarray}
\chi + N (e) \to \nu + N (e)\;,
\end{eqnarray}
where $\chi$ denotes the DM particle.
In particular, the energy conservation of the initial and final states in this process induces a localized recoil energy of nucleus or electron
\begin{eqnarray}
E_R=m_\chi^2/2m_{N,e}\;.
\label{eq:ER}
\end{eqnarray}
This kinematics exhibits a distinct peak-like signature in the scattering rate~\cite{Dror:2019onn,Dror:2019dib}, rather than the smooth distribution in ordinary elastic scattering.
The possible peak-like energy deposit would enable us to search for the relevant interaction between DM and neutrino in this way. In fact, the inverse process, i.e., the conversion from neutrino to an exotic fermion~\cite{Ge:2020jfn,Shoemaker:2020kji,Shakeri:2020wvk,Hurtado:2020vlj,Chen:2021uuw}, has been utilized to explain the recent XENON1T excess~\cite{XENON:2020rca}. It turns out that the neutrino magnetic dipole portal can account for the excess of keV electron recoil events because the induced scattering rate is inversely proportional to the recoil energy~\cite{Miranda:2020kwy,Babu:2020ivd,Shoemaker:2020kji,Brdar:2020quo,AristizabalSierra:2020zod}. It is thus natural to consider the above fermionic DM absorption through the transition from dark magnetic dipole and the dynamical conversion to neutrino.
Naively speaking, according to the recoil energy formula in Eq.~(\ref{eq:ER}), the absorption by the nucleus and electron targets in direct detection experiments is sensitive to sub-GeV or sub-MeV DM, respectively.

In this work we consider a Dirac fermion DM $\chi$ charged under a dark gauge group $U(1)'$ and the dark magnetic dipole operator. The dark gauge field $A'$ interacts with the Standard Model (SM) electromagnetic field through a kinetic mixing term. The DM field mixes with right-handed neutrino and the DM$-\nu$ tensor current interacts with the ordinary electromagnetic charge current via the kinetic mixing term of gauge fields. As a result, the incoming DM is absorbed and converted into neutrino in final states intermediated by dark photon through the dipole-charge interaction. We also estimate the contribution from the nuclear magnetic dipole by evaluating the nuclear magnetic dipole form factor for inelastic DM-nucleus scattering. The inelastic scattering off electron target will also be studied by calculating the ionization form factor of the bound electrons.
We aim to obtain the prospective sensitivity of direct detection experiments to the dark magnetic dipole moment, the kinetic mixing of gauge fields as well as the DM$-\nu$ coupling.

This paper is organized as follows. In Sec.~\ref{sec:model} we describe the dark photon model with DM magnetic dipole moment. In Sec.~\ref{sec:nucleus} we study the DM absorption by nuclear targets, and display the projected constraint on the total cross section as well as the couplings. The ionization signal in the DM absorption by electron target will be discussed in Sec.~\ref{sec:electron}. Our main conclusions are summarized in Sec.~\ref{sec:Con}. We show how we calculate the nuclear magnetic dipole form factor and the ionization form factor in Appendices.

\section{The dark photon model with magnetic dipole moment}
\label{sec:model}

\begin{figure}
\centering
\includegraphics[scale=0.8,width=0.4\linewidth]{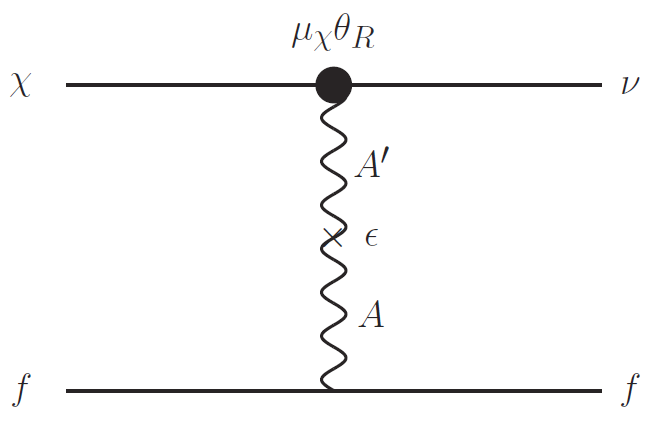}
\includegraphics[scale=0.8,width=0.4\linewidth]{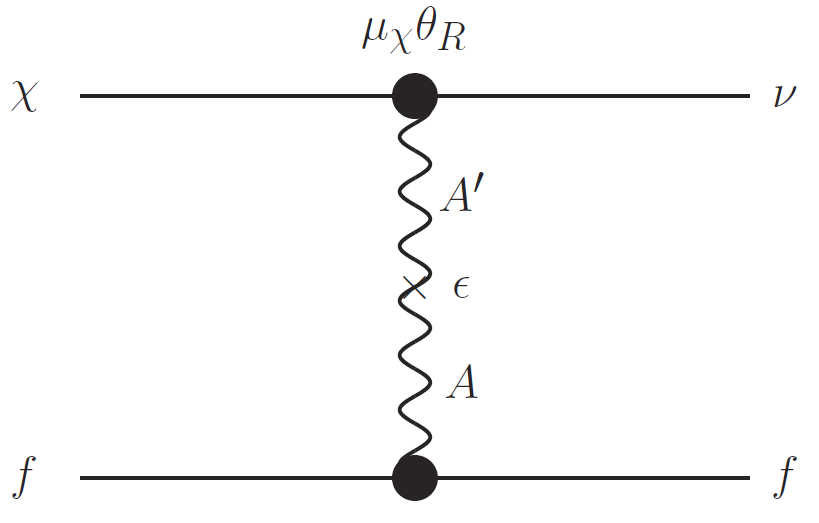}
\caption{Diagrams for DM absorption from dipole-charge (left) or dipole-dipole (right) interaction in our model. The black dot denotes the magnetic dipole operator.
}
\label{fig:diagram}
\end{figure}

We consider a Dirac fermion $\chi$ charged under a dark gauge group $U(1)'$. The magnetic dipole operator of $\chi$ and the gauge Lagrangian of the dark gauge field $A'_\mu$ are given by
\begin{eqnarray}
\mathcal{L}\supset {\mu_\chi\over 2}\bar{\chi}\sigma^{\mu\nu}\chi F'_{\mu\nu}+{1\over 4}F^{\prime\mu\nu}F'_{\mu\nu}+{\epsilon\over 2}F^{\mu\nu}F'_{\mu\nu}+{1\over 2}m_{A'}^2 A^{\prime\mu}A'_\mu\;,
\label{eq:Lgauge}
\end{eqnarray}
where $\mu_\chi$ is the dipole strength, $F_{\mu\nu}=\partial_\mu A_\nu - \partial_\nu A_\mu$ is the SM electromagnetic field tensor and $F'_{\mu\nu}=\partial_\mu A'_\nu - \partial_\nu A'_\mu$ is the field strength tensor of $U(1)'$. For dark photon, see Ref.~\cite{Fabbrichesi:2020wbt} for a recent review and references therein.
A scalar field $\phi$ charged under $U(1)'$ is then introduced and the relevant Lagrangian is
\begin{eqnarray}
\mathcal{L}\supset m_\chi \bar{\chi}\chi + y\phi \bar{\chi}P_R\nu + h.c.\;.
\end{eqnarray}
The right-handed fields $\chi_R$ and $\nu_R$ are then mixed and shifted in terms of a mixing angle $\theta_R=y\langle\phi\rangle/m_\chi$~\cite{Dror:2019onn,Dror:2019dib,Dror:2020czw}. Thus, the $\chi_R$ eigenstate has an additional component $-\theta_R \nu_R$.
Suppose the SM fermions are not charged under $U(1)'$, the diagonalization of the kinetic mixing term in Eq.~(\ref{eq:Lgauge}) gives the coupling of dark photon $A'$ to the ordinary electromagnetic current $-e \epsilon A'_\mu J^\mu_{em}$.
Then we have the following effective interaction as shown in the left panel of Fig.~\ref{fig:diagram}
\begin{eqnarray}
\mathcal{L}_{eff}&\supset&{\epsilon eQ_f \mu_\chi \theta_R\over q^2-m_{A'}^2}q^\mu\overline{\chi_L}\sigma_{\mu\nu}\nu_R \bar{f}\gamma^\nu f+h.c.\;,
\label{eq:Neffint_dipch}
\end{eqnarray}
where $q$ is the momentum transfer carried by the dark photon, $f$ denotes the SM quark or electron and $Q_f$ is the electromagnetic charge of fermion $f$.
One can see that the DM only scatters off proton or electron but not neutron as the electromagnetic charge of neutron is zero.

On the other hand, some nuclei targets have sizable magnetic dipoles such as cesium (Cs), iodine (I) or a couple of xenon's isotopes. Thus, besides the above dipole-charge (DC) interaction, there could be dipole-dipole (DD) contribution in the scattering on a nucleus when the dark photon interacts with the nuclear magnetic dipole as shown in the right panel of Fig.~\ref{fig:diagram}. The scattering induced by the dipole-dipole interaction may dominate over the dipole-charge scattering and accounts for the DAMA annual modulation signature~\cite{Chang:2010en,Barger:2010gv}.
The dipole-dipole interaction at the nuclear level in our model is given by
\begin{eqnarray}
{\epsilon \mu_\chi \mu_N\theta_R\over q^2-m_{A'}^2}q^\mu q^\alpha g^{\nu\beta}\overline{\chi_L}\sigma_{\mu\nu}\nu_R \bar{N}\sigma_{\alpha\beta} N+h.c.\;,
\end{eqnarray}
where $\mu_N$ denotes the nuclear magnetic moment.

To guarantee that the fermion $\chi$ serves as a stable DM particle, the lifetime of $\chi$ should be longer than the age of the Universe, i.e. $t_{\rm Universe}=4.4\times 10^{17}~{\rm sec}$~\cite{Ade:2015xua}.
Requiring the DM being stable at the Universe time scale would set a stringent bound on the coupling and the mass of DM particle.
For the above model mediated by dark photon, the leading decay channel would be $\chi\to \nu\gamma\gamma\gamma$ for light DM~\cite{Dror:2020czw}. The two-body decay $\chi\to \nu A'$ can be forbidden by requiring $m_\chi<m_{A'}$.
We define $U=\mu_\chi \epsilon \theta_R$ and recall the decay rate in Refs.~\cite{Dror:2019onn,Dror:2019dib,Dror:2020czw} as
\begin{eqnarray}
\Gamma(\chi\to \nu\gamma\gamma\gamma)\simeq 10^{-32}~{\rm s}^{-1}~\Big({m_\chi\over 100~{\rm keV}}\Big)^{15}~\Big({\rm TeV}\cdot e U\Big)^2~\Big({{\rm GeV}\over m_{A'}}\Big)^4\;.
\label{eq:con}
\end{eqnarray}
Note that this result assumes that the dark photon mass is larger than the momentum transfer.

\section{Dark Matter Absorption by Nuclear Targets}
\label{sec:nucleus}

For DM absorption by nuclear targets, i.e., $\chi + N \to \nu + N$, the differential scattering cross section induced by neutral current (NC) is given by
\begin{eqnarray}
{d\sigma_{NC}\over dE_R}&=&{d\sigma_{DC}\over dE_R}+{d\sigma_{DD}\over dE_R}\nonumber\\
&=&{\alpha \mu_\chi^2 \theta^2_R \epsilon^2 m_\chi\over v m_N (m_{A'}^2+2m_N E_R)^2} Z^2 F_Z^2(E_R) \delta(E_R-E_R^0)\nonumber \\
&&\Big(6m_N^2 m_\chi E_R - 8m_N^2 E_R^2 -m_N m_\chi^3+4m_N m_\chi^2 E_R -2 m_N m_\chi E_R^2+m_\chi^3 E_R \Big)\nonumber\\
&+&{\mu_\chi^2 \mu_{N}^2 \theta^2_R \epsilon^2 m_\chi E_R \over 2\pi v (m_{A'}^2+2m_N E_R)^2}{I+1\over 3I} F_D^2(E_R) \delta(E_R-E_R^0)\nonumber \\
&&\Big(-4m_N^2 E_R^2+8m_N m_\chi^2 E_R-8m_N m_\chi E_R^2+2m_N E_R^3+m_\chi^4+4m_\chi^3 E_R-3m_\chi^2 E_R^2\Big) \;,\nonumber \\
\label{eq:Ndiffxsec}
\end{eqnarray}
where $E_R^0=m_\chi^2/2m_N$, $v$ is the DM velocity, $Z$ is the atomic number, $I$ is the nuclear spin, $F_Z(E_R)$ is the ordinary nuclear form factor and $F_D(E_R)$ is the nuclear magnetic dipole form factor. The two terms in the above differential cross section correspond to the dipole-charge (DC) interaction and the dipole-dipole (DD) interaction, respectively. The kinematics in this inelastic scattering with nearly massless neutrino in final states simplifies the energy conservation and leads to the delta functions in Eq.~(\ref{eq:Ndiffxsec}) by only keeping the $\mathcal{O}(v^0)$ terms. As a result, the fermion DM absorption recoil is peaked at $m_\chi^2/2m_N$. This is the key difference between the DM absorption and the usual elastic scattering.

\begin{figure}
\centering
\includegraphics[scale=0.8,width=0.8\linewidth]{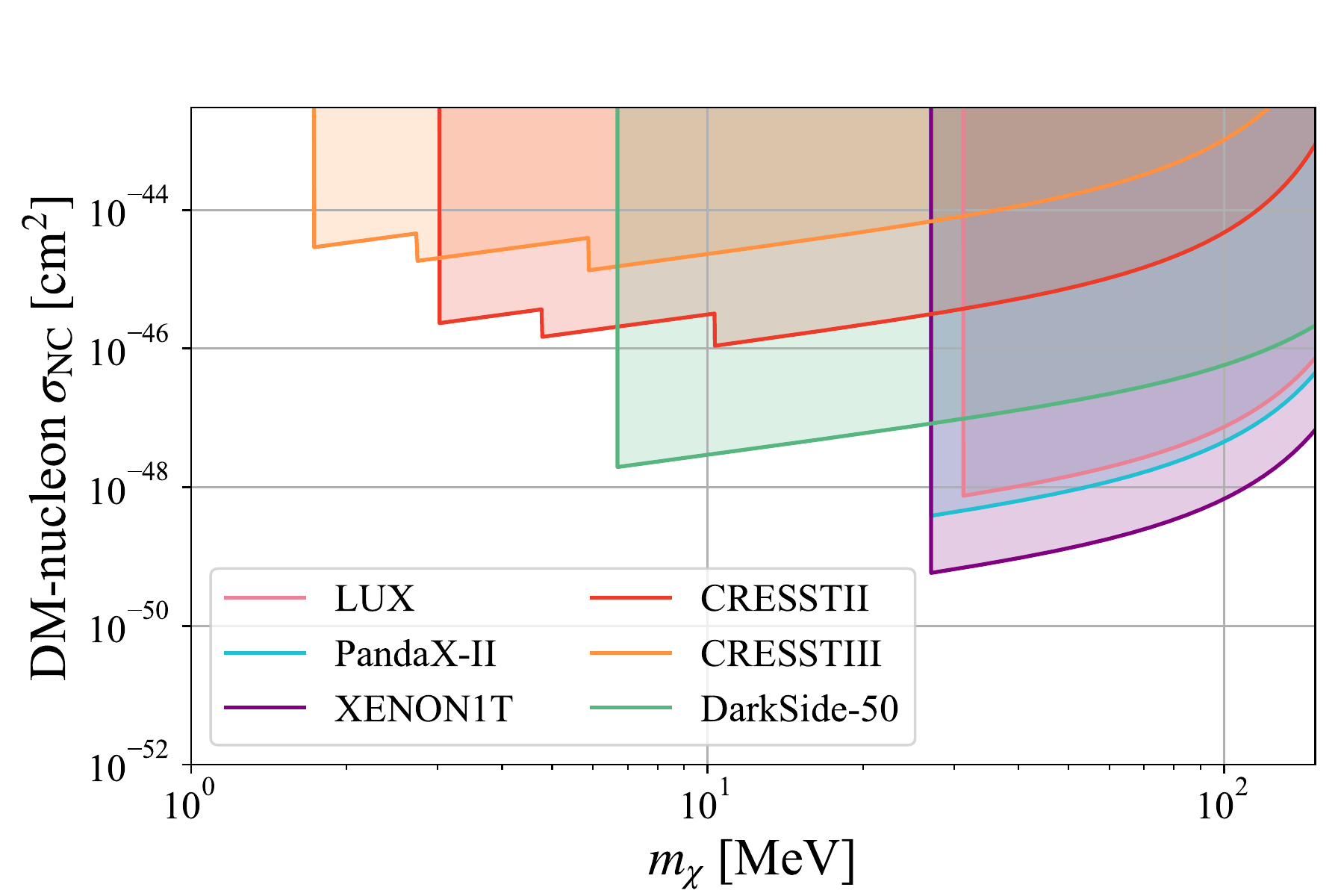}
\caption{Projected constraints on the DM-nucleon cross section $\sigma_{NC}$ in our model as a function of $m_\chi$. The constraints are given for different experiments: LUX (pink)~\cite{LUX:2016ggv}, PandaX-II (blue)~\cite{PandaX-II:2017hlx}, XENON1T (purple)~\cite{XENON:2018voc}, CRESSTII (red)~\cite{CRESST:2015txj}, CRESSTIII (orange)~\cite{CRESST:2017cdd} and DarkSide-50 (green)~\cite{DarkSide:2014llq,DarkSide:2018bpj}.
The target isotopes include $\rm^{131}{Xe}$ for LUX, PandaX-II, and XENON1T, $\rm^{186}{W}, \rm^{40}{Ca}, \rm^{16}{O}$ for CRESSTII and CRESSTIII, and $\rm^{40}{Ar}$ for DarkSide-50.
}
\label{fig:sigmaNC}
\end{figure}

Given the delta function in Eq.~(\ref{eq:Ndiffxsec}), the DM-nucleus scattering rate from the dipole-charge interaction can be easily obtained as~\cite{Dror:2019dib}
\begin{eqnarray}
\label{equ:12}
R=\frac{\rho_\chi}{m_\chi}\sigma_{NC}Z^2\sum_j{N_{T,j}F_j(q)^2\Theta(E_{R,j}^0-E_{th})}\;,
\label{eq:Nrate}
\end{eqnarray}
where $\rho_\chi$ is the local DM density, $\sigma_{NC}$ is the total cross section of inelastic DM-nucleon scattering, $N_T$ and $F(q)$ correspond to the target number per detector mass and nuclear form factor, respectively, and $\Theta$ is a step function enforcing the minimal recoil energy to be the energy threshold $E_{th}$ of the detector.
All isotope targets in the experiments should be summed over the index $j$.
Note that we replace the atom mass $A$ in the Eq.~(3.8) of Ref.~\cite{Dror:2019dib} by the atomic number $Z$ here for our model and extract it out of Eq.~(\ref{eq:Ndiffxsec}). Moreover, as we drop the high order terms of DM velocity $O(v)$, the integration over DM velocity distribution $f(v)$ should be normalized, i.e., $\int{d^3v f(v)}=1$.
When only considering the dipole-charge contribution in the first term of Eq.~(\ref{eq:Ndiffxsec}),
one can project a bound on $\sigma_{NC}$ by requiring a certain number of absorption events in any given experiment (here for simplicity $<10$ events with the exposures and $E_{th}$ values in Table~3 of Ref.~\cite{Dror:2019dib}).
In Fig.~\ref{fig:sigmaNC} we show the projected constraints on the DM-nucleon cross section $\sigma_{NC}$ in our model as a function of $m_\chi$. The lower limit of DM mass is determined by the recoil energy threshold in each experiment. For instance, XENON1T can probe the DM mass above 27 MeV and is sensitive to $\sigma_{NC}\gtrsim 10^{-49}$ cm$^2$.
The kinks occur for different DM masses due to the atomic numbers of different target isotopes and the step function in Eq.~(\ref{eq:Nrate}) determines the abscissa of kink according to $m_\chi=\sqrt{2m_N E_{th}}$.
In addition, we consider the DM mass less than 150 MeV because the nuclear recoil energy threshold is normally smaller than 100 keV. In this range, the influence caused by nuclear form factor can be neglected.

\begin{figure}[htb!]
\centering
\subfigtopskip=-1pt
\subfigbottomskip=2pt
\subfigcapskip=-5pt
\subfigure[~$m_\chi$ = 10 MeV]{
\includegraphics[width=0.45\linewidth]{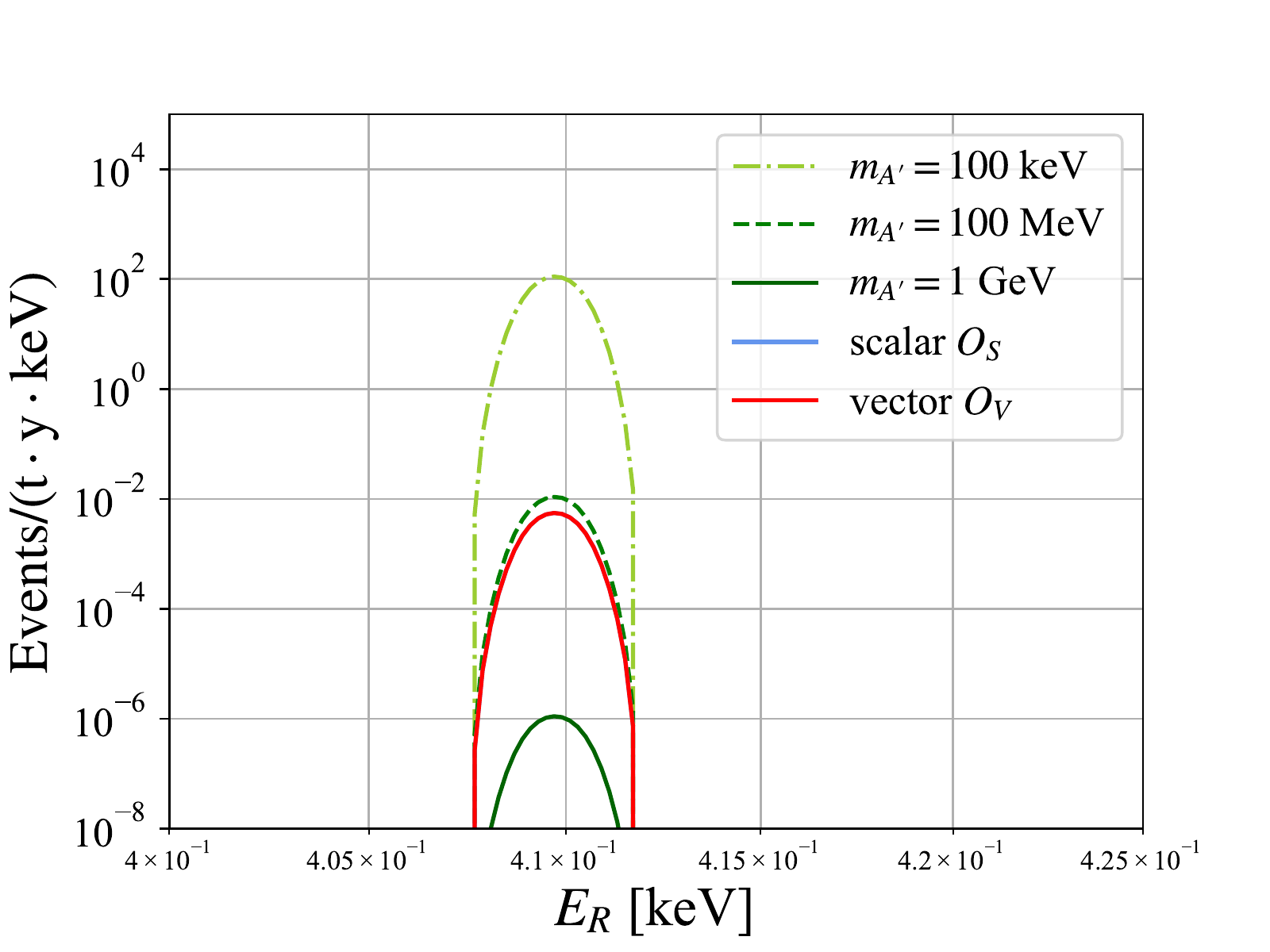}}
\subfigure[~$m_\chi$ = 100 MeV]{
\includegraphics[width=0.45\linewidth]{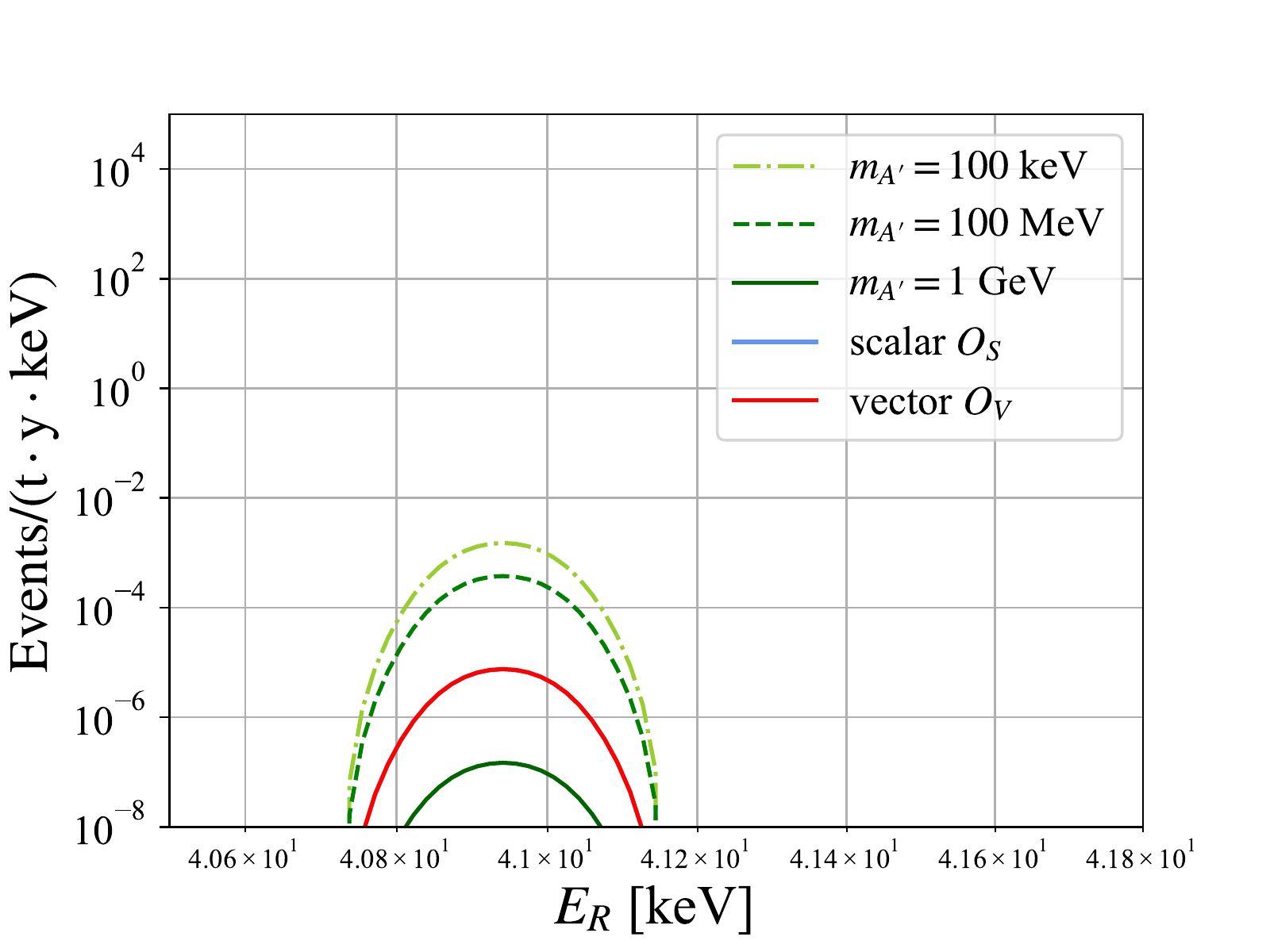}}
\caption{Differential scattering rates of target $\rm{^{131}Xe}$ for some masses of DM and dark photon. The coupling $e^2U^2$ in Eq.~(\ref{eq:RDCdiff}) is fixed to be $10^{-50}~\rm{cm^2}$. The differential rates of the scalar and vector interactions with $1/\Lambda^2=10^{-50}~\rm{cm^2}$ in Ref.~\cite{Dror:2019onn} are also shown for reference.
}
\label{fig:Ndiff}
\end{figure}

Note that the above constraint on $\sigma_{NC}$ given by Eq.~(\ref{eq:Nrate}) is independent of the DM absorption models. Next we wonder the specific constraint on the effective interaction in Eq.~(\ref{eq:Neffint_dipch}) and take into account higher order expansion of DM velocity. After integrating Eq.~(\ref{eq:Nrate}) over the DM velocity distribution $f(v)$, the differential scattering rate per nuclear recoil energy for the dipole-charge interaction is given by
\begin{eqnarray}
\frac{dR_{DC}}{dE_R}&=&N_T \frac{\rho_\chi}{m_\chi}\sqrt{\frac{E_R}{2m_N}}\frac{e^2 U^2 m_N}{4\pi m_\chi p_\nu(m_{A'}^2+2m_N E_R)^2}Z^2F_Z^2(E_R)\int{d^3\boldsymbol v\frac{f(\boldsymbol v)}{v}\Theta(v-v_{\rm{min}})}\nonumber\\
&\times&[6m_N^2 m_\chi E_R-8m_N^2 E_R^2-m_N m_\chi^3+4m_N m_\chi^2 E_R-2m_N m_\chi E_R^2+m_\chi^3 E_R] \;,
\label{eq:RDCdiff}
\end{eqnarray}
where $p_\nu$ is neutrino momentum and $U$ is defined as $U = \mu_\chi\theta_R\epsilon$. Since the velocity distribution is non-trivial in this case, we use the truncated Maxwell distribution $f(v)$~\cite{Jungman:1995df,Belanger:2008sj}. The $f(v)$ is truncated at $\Theta(v_{\rm{esc}}-\vert \vec{v}+\vec{v}_e(t)\vert)$, according to the DM standard halo model where $\vec v$ is the speed of DM, $v_{\rm{esc}}\sim 550~\rm{km/s}$ is the galactic escape velocity, and $\vec{v}_e(t)$ is the speed of earth which can be decomposed into the Sun's motion in the Galaxy and the Earth's motion in solar system orbit ($\vec{v}_\odot(t) + \vec{v}_\oplus(t)$). In Fig.~\ref{fig:Ndiff} we show the differential scattering rate of target $\rm{^{131}Xe}$ for some masses of DM and dark photon. One can see that the distributions indeed peak around $m_\chi^2/2m_N$. Although lighter DM produces more events, the recoil energy of these nuclear scatterings is less than $1~{\rm keV}$ when $m_\chi < 10~{\rm MeV}$ which is too small to be detected by current detectors. For the majority of DM detection experiments, the recoil energy thresholds are generally greater than $1~{\rm keV}$. Furthermore, the widths of the spectra in Fig.~\ref{fig:Ndiff} are also very small. Different target isotopes result in a tiny distinction for the spectra. For example, in CRESSTII the target $\rm{CaWO_4}$ contains four isotopes of tungsten: $^{182}{\rm W}$, $^{183}{\rm W}$, $^{184}{\rm W}$ and $^{186}{\rm W}$~\cite{CRESST:2015txj} which are only distinguishable when the energy resolution is less than $50~{\rm eV}$. For comparison, by taking XENON1T experiment for illustration, the resolution of the reconstructed energy yields 5\% (1.4\%) for energy deposit 41.5 (609) keV as shown in Ref.~\cite{XENON:2020iwh}. Thus, the typical energy resolution there is about 2 keV for 40 keV recoil energy.

\begin{figure}[htb!]
\subfigcapskip=2 pt
\hspace{-0.2 cm}
\begin{minipage}[t]{0.46\textwidth}
\centering
\includegraphics[scale=0.5]{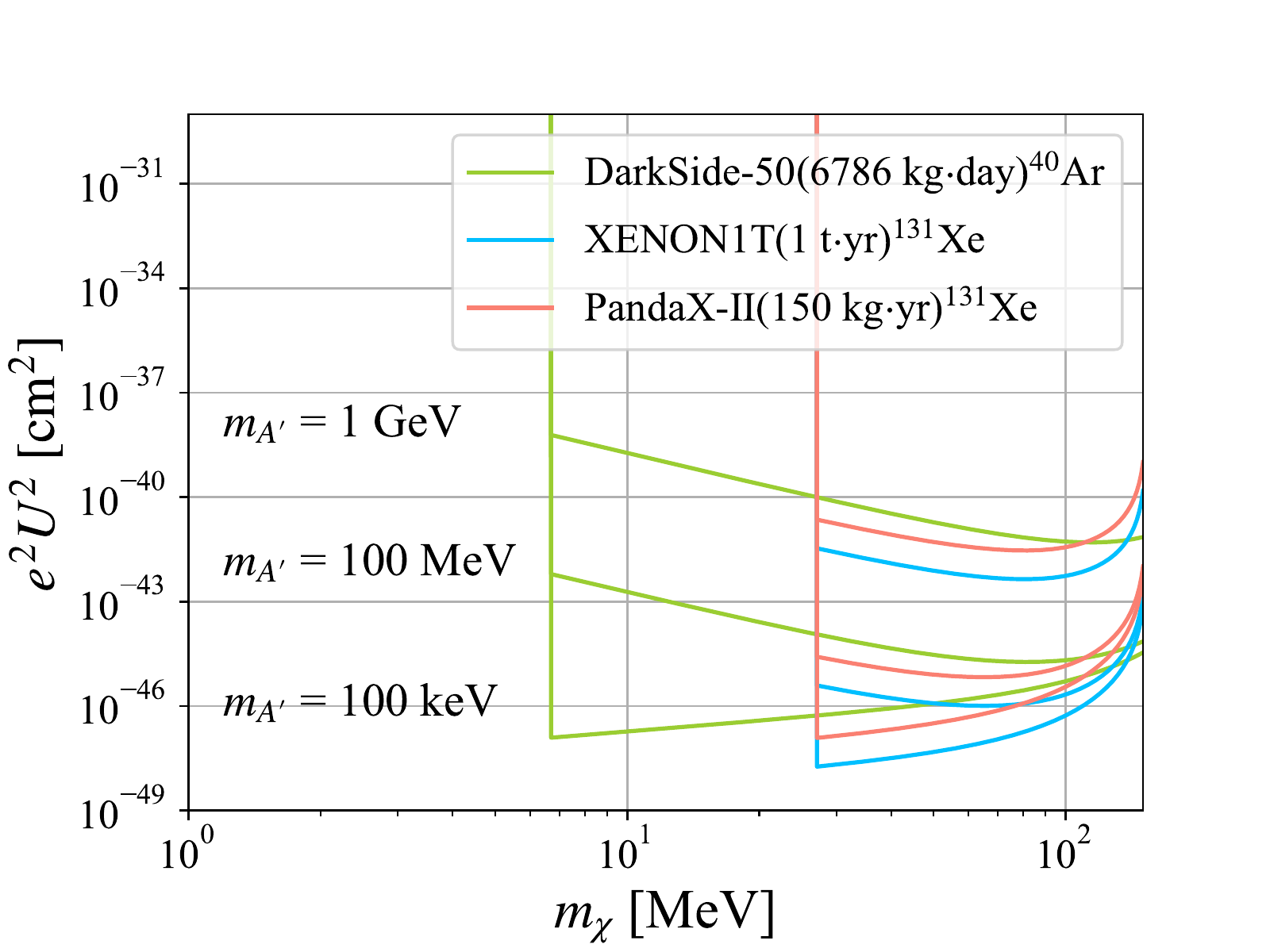}
\end{minipage}
\begin{minipage}[t]{0.65\textwidth}
\centering
\includegraphics[scale=0.5]{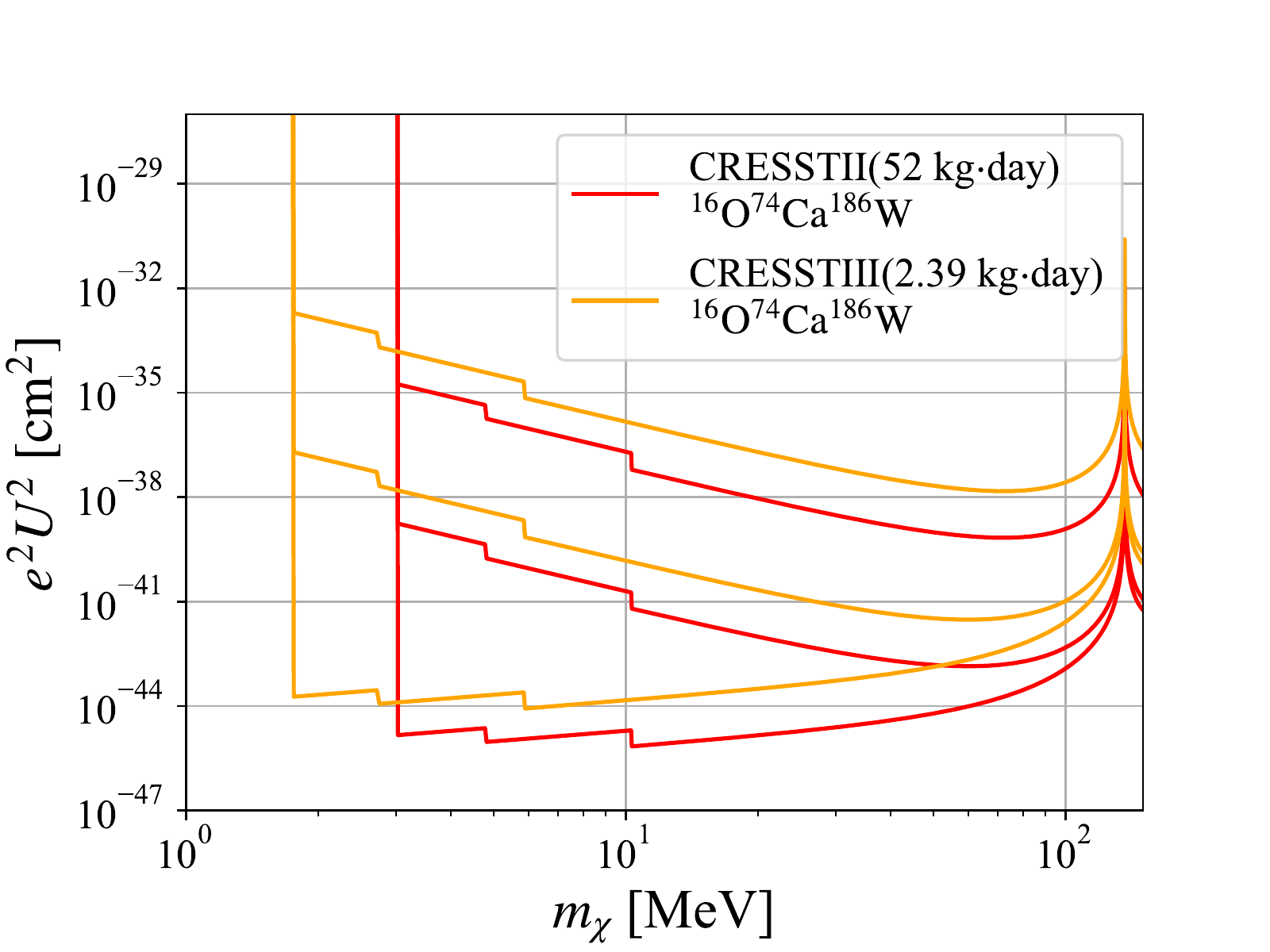}
\end{minipage}
\caption{The projected constraint on $e^2U^2$ from dipole-charge contribution for both liquid (left) and crystal (right) targets, assuming different dark photon masses. }
\label{fig:NeU}
\end{figure}

To constrain the coupling $e^2U^2$, after integrating over the recoil energy, we have the total scattering rate due to the dipole-charge interaction
\begin{eqnarray}
R_{DC}=N_T\left(\frac{\rho_\chi}{m_\chi}\right)\frac{e^2U^2}{4\pi(m_{A'}^2+m_\chi^2)^2}\left(2m_\chi^4\right)Z^2F_Z^2(q)\Theta(E_R^0-E_{th})\;,
\label{eq:NRDC}
\end{eqnarray}
where $q=|\textbf{q}|=\sqrt{2m_N E_R}\sim m_\chi$.
The only unknown parameter is the dark photon mass $m_{A'}$. In Fig.~\ref{fig:NeU} we show the projected constraint on $e^2U^2$ for both liquid (left) and crystal (right) targets, assuming different dark photon masses.
Lower $e^2U^2$ can be constrained for decreasing $m_{A'}$, for instance $e^2U^2\simeq 10^{-47}$ cm$^2$ or $10^{-45}$ cm$^2$ for $m_{A'}=100$ keV. The target isotope of DarkSide-50 is $^{40}\rm{Ar}$ which is lighter than $^{131}\rm{Xe}$ and its recoil energy threshold is 50 times lower than those of XENON1T and PandaX-II. Thus, as shown in the left panel, the DarkSide-50 bounds can reach a lower DM mass limit. The target isotope of CRESSTII and CRESSTIII is $\rm{CaWO}_4$~\cite{CRESST:2015txj,CRESST:2017cdd}. As seen from the right panel, due to the threshold $\Theta (E_R^0-E_{th})$, there appear three kinks corresponding to oxygen ($^{16}\rm{O}$), calcium ($^{40}\rm{Ca}$) and tungsten ($^{186}\rm{W}$) from left to right.
For large DM mass, there is an apparent increasing behavior due to the strong suppression of Helm form factor $F_Z(q)$ with high momentum transfer.

For the nuclei targets discussed above, there is a lower limit of $m_\chi\sim {\rm MeV}$ for the searching capability due to the detector energy threshold at keV level and the nuclear masses. There exist proposals of future experiments with light nuclei (such as Hydrogen or Lithium~\cite{Budnik:2017sbu,Szydagis:2018wjp}) or semiconductor/superconductor (see Ref.~\cite{Essig:2022dfa} for a recent review) and low-threshold at the level of 1 eV or even less. As a result, the detectable DM mass can reach as low as $\mathcal{O}(0.01)$ MeV~\cite{Dror:2019dib}. To explore sub-MeV DM mass region, we also consider new scattering strategy which will be discussed in next section.

Moreover, it turns out that the DM lifetime constraint becomes quite severe for the sensitive region of the above DM-nucleus scattering.
For the DM mass range of $m_\chi\gtrsim 1~{\rm MeV}\gtrsim 2m_e$, the dominant decay channel would be $\chi\to \nu ee$ at tree-level. The decay width becomes $\Gamma(\chi\to \nu ee)\simeq 10^{-8}~{\rm s}^{-1}~\Big({m_\chi\over 1~{\rm MeV}}\Big)^{7}~\Big({\rm TeV}\cdot e U\Big)^2~\Big({{\rm GeV}\over m_{A'}}\Big)^4$. For lower DM masses the dominant decay is $\chi\to \nu \gamma\gamma\gamma$ given in Eq.~(\ref{eq:con}). These constraints rule out the sensitive region of DM-nucleus scattering in Fig.~\ref{fig:NeU} for those nuclei targets. %
To avoid such constraints, in theoretical aspect, one needs advanced model building and induce necessary fine-tuning in the UV completion. In an alternative model, we can consider SM fermions charged under the new $U(1)'$ and the charge current is $J'_\mu=Q'_fg'\bar{f}\gamma_\mu f$ but with non-universal charges $Q'_f$. Thus, the diagrams in above decay processes are induced by $A'$ propagator without additional kinetic mixing. We further assume leptophobic $A'$ to forbid $\chi\to\nu ee$ decay. Moreover, the charges of SM quarks are fine-tuned in the loop of $\chi\to \nu \gamma\gamma\gamma$ process. As a result, the decay widths are highly suppressed. Nevertheless, the UV completion would be quite dedicated to avoid the lifetime constraint.
In the aspect of experiment, as mentioned above, future low-threshold detectors would have energy threshold at eV level or even less and be sensitive to much lower DM masses~\cite{Dror:2019dib}. In addition, other detection strategies such as cosmic-ray boosted DM~\cite{Bringmann:2018cvk} or Midgal effect~\cite{Ibe:2017yqa} would also lower the detectable DM mass range. These improvements can also help to evade the lifetime constraint. We leave the relevant studies to a future work. This also motivates us to explore lighter DM through the scattering of DM and bound electrons.

\begin{figure}
\hspace{-0.2 cm}
\begin{minipage}[t]{0.46\textwidth}
\centering
\includegraphics[scale=0.5]{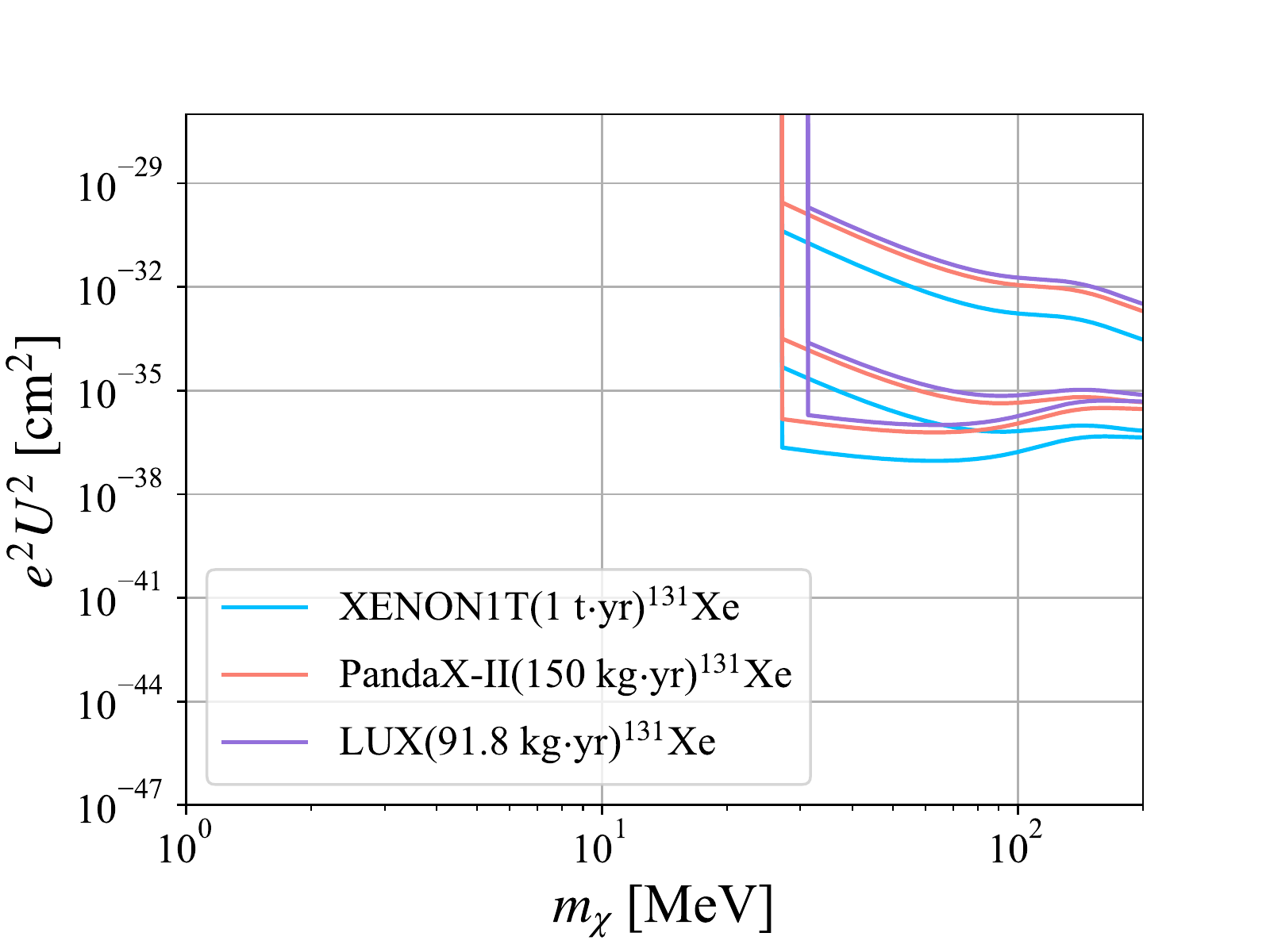}
\end{minipage}
\begin{minipage}[t]{0.65\textwidth}
\centering
\includegraphics[scale=0.5]{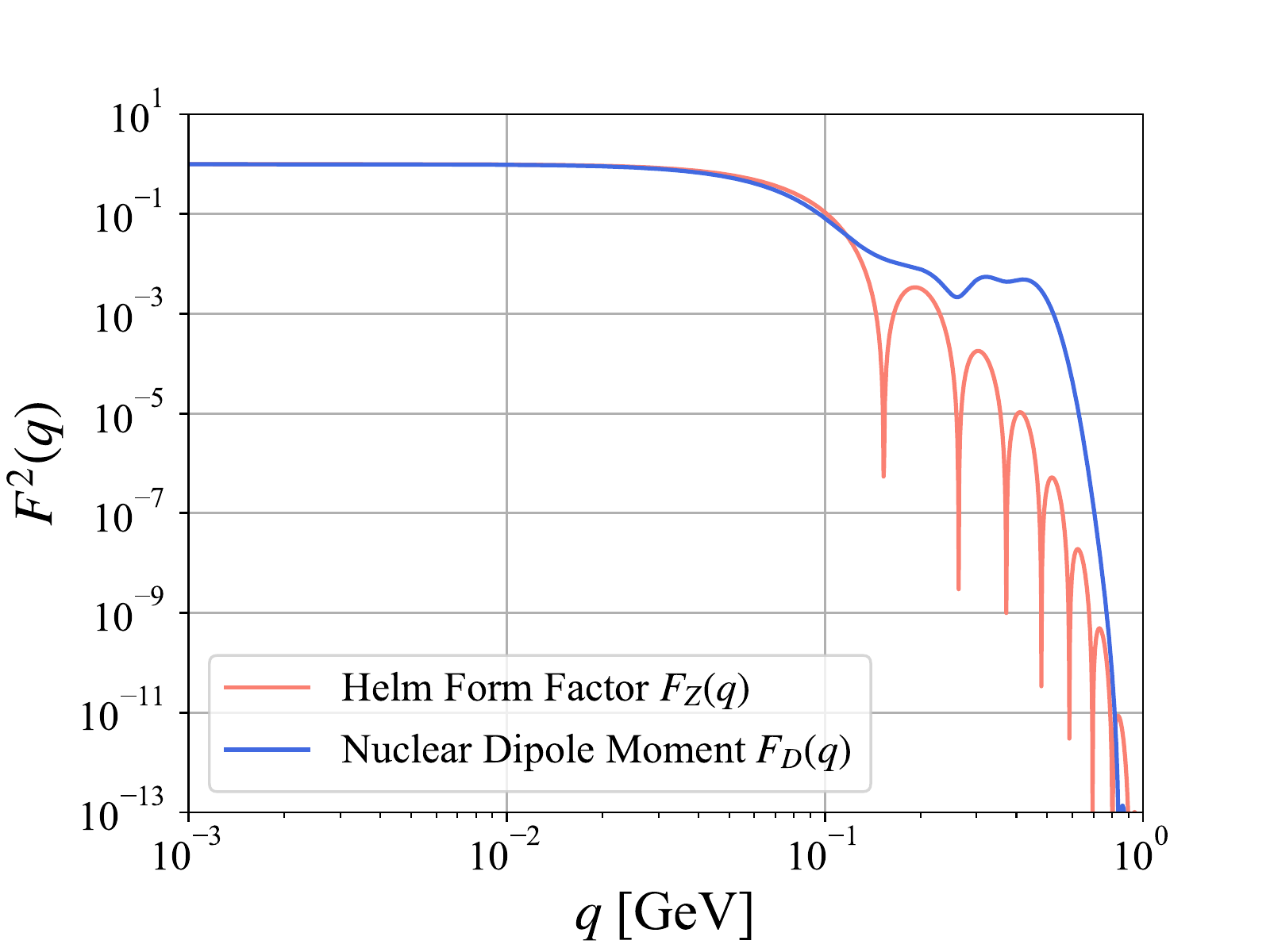}
\end{minipage}
\caption{Left: The projected constraint on $e^2U^2$ from dipole-dipole contribution for $^{131}\rm{Xe}$ target. The assumed dark photon masses are the same as those in Fig.~\ref{eq:NRDC}. Right: The comparison of form factors $F_Z^2(q)$ and $F_D^2(q)$.}
\label{fig:NeUDD}
\end{figure}

In addition, we can also take into account the dipole-dipole contribution in Eq.~(\ref{eq:Ndiffxsec}) and obtain the relevant scattering rate as follows
\begin{eqnarray}
R_{DD} = N_T\left(\frac{\rho_\chi}{m_\chi}\right)\frac{\mu_N^2U^2}{2\pi(m_{A'}^2+m_\chi^2)^2}\frac{I+1}{3I}\left(\frac{m_\chi^3}{2m_N}\right)F_D^2(q)\frac{m_\chi^4}{2m_N^2}(8m_N^2-m_\chi^2)\Theta(E_R^0-E_{th})\;,
\nonumber \\
\label{eq:NRDD}
\end{eqnarray}
where the nuclear magnetic dipole form factor $F_D$ includes the contributions from both angular momentum and nuclear spin. Recall that in DAMA experiment (target \rm{NaI}) iodine has large nuclear magnetic moment and large mass, thus it is necessary to consider the dipole-dipole contribution. Although the magnetic dipole moment of xenon is not as large as iodine, we wonder here whether the dipole-dipole contribution from xenon target can induce sizable constraint for the given scattering process in Eq.~(\ref{eq:Ndiffxsec}). According to Appendix~\ref{app:FD}, the form factor $F_D$ for $^{131}\rm{Xe}$ can be written as
\begin{eqnarray}
F_D^2(q)=\left(0.4\frac{L(q)}{L(0)}+0.6\sqrt{\frac{S(q)}{S(0)}}~\right)^2\;,
\end{eqnarray}
where $L(q)$ (the spin-independent Helm form factor) and $S(q)$ are the angular momentum and nuclear spin contributions to the magnetic dipole moment, respectively. The detailed expression of $S(q)$ is given by polynomial fitting and the fitted coefficient can be found in Ref.~\cite{Ressell:1997kx}. For $^{131}\rm{Xe}$ one has $I=3/2$ and $\mu_N=0.692e/2m_p$~\cite{Barger:2010gv}.
We then display the bound on $e^2U^2$ from Eq.~(\ref{eq:NRDD}) for $^{131}\rm{Xe}$ in the left panel of Fig.~\ref{fig:NeUDD}. It shows that there is no significant contribution from dipole-dipole interaction for the inelastic DM absorption. Note that, since the $O(v)$ terms are neglected, we here make an approximation of the momentum transfer that is $q=\sqrt{E_R^2+2m_NE_R}\approx m_\chi$ with $E_R \approx m_\chi^2/2m_N$. Thus, the form factors $F_D(q)$ and $F_Z(q)$ are more precise for low DM masses $m_\chi\ll 1$ GeV as shown in the right panel of Fig.~\ref{fig:NeUDD}.

\section{Dark Matter Absorption by Electron Targets}
\label{sec:electron}

As seen from the above section, the DM absorption by nuclear targets leads to distinctive peak-like scattering signal. However, due to the limited energy resolution, it is very likely to suffer from the difficulties of identifying the energy deposition and distinguishing the target isotopes.
The fermionic DM can also be absorbed by bound electron targets, which induces ionization signal.
Such signal would be sensitive to lower DM mass and the electron binding energies from both inner and outer atomic shells make the localized recoil energy $E_R=m_\chi^2/2m_e$ spread out.

According to the effective Lagrangian in Eq.~(\ref{eq:Neffint_dipch}), induced by the dipole-charge interaction, we have the differential scattering cross section of DM absorption by electron
\begin{eqnarray}
\frac{d\langle\sigma_{\rm{ion}}^{nl}v\rangle_{DC}}{dE_R}&=&\frac{\overline{\vert\mathcal M\vert^2}}{64\pi m_\chi m_e^2}\frac{q}{E_R}\vert f_{\rm{ion}}^{nl}(k',q)\vert^2\Theta(q)\nonumber\\
&=&\frac{e^2U^2}{16\pi(m_{A'}^2-m_\chi(m_\chi-2q))^2}\frac{q}{E_R}\frac{m_\chi^2}{m_e^2}\vert f_{\rm{ion}}^{nl}(k',q)\vert^2\Theta(q)\nonumber \\
&&\times[-4m_em_\chi(m_e+m_\chi)-m_\chi^3+3(m_\chi^2+2m_e^2+4m_em_\chi)q-2(m_\chi+4m_e)q^2]\;,\nonumber \\
\label{eq:ediffxsec}
\end{eqnarray}
where the bracket $\left<\cdots\right>$ denotes the integral of the Maxwell-Boltzmann DM velocity distribution $f(v)$ as mentioned in Sec.~\ref{sec:nucleus}, and $k'=\sqrt{2m_e E_R}$ is the momentum of the electron in the shell $(n,l)$.
For the electron scattering, we should sum over all possible shells together and the differential scattering rate is
\begin{eqnarray}
\frac{dR_{\rm{ion}}}{dE_R}=N_T\frac{\rho_\chi}{m_\chi}\sum_{nl}\frac{d\langle\sigma_{\rm{ion}}^{nl}v\rangle_{DC}}{dE_R}\;.
\end{eqnarray}
To evaluate the integral of the scattering cross section, one should specify the electron ionization form factor $|f_{\rm{ion}}^{nl}(k',q)|$ which becomes
an important part in the calculation of the fermionic DM absorption by electron target.

The general form of the ionization form factor $|f_{\rm{ion}}^{nl}(k',q)|$ can be given in terms of Wigner 3-$j$ symbol, spherical Bessel functions $j_L$ and radial wave-functions~\cite{2012RE,2017RE}
\begin{equation}
|f_{\rm{ion}}^{nl}(k',q)|^2 = \frac{4k'^3}{(2\pi)^3}\sum_{l'L}(2l+1)(2l'+1)(2L+1)
\times \left[
\begin{matrix}
l~~l'~~L\\
0~~0~~0
\end{matrix}
\right]^2 \left|\int{r^2drR_{k'l'}(r)R_{nl}(r)j_L(qr)}\right|^2\;,
\label{eq:ionFF}
\end{equation}
where the sum $\sum_{l'L}$ denotes $\sum_{l'=0}^\infty\sum_{L=\vert l-l'\vert}^{l+l'}$ (excluding $L=0$).
Here we have used the following fact that the angular integral of the product of three spherical harmonics can be given by~\cite{ARFKEN2013773}
\begin{equation}
\int{d\Omega Y_{l_1}^{m_1}(\theta,\phi)Y_{l_2}^{m_2}(\theta,\phi)Y_{l_3}^{m_3}(\theta,\phi)}=\sqrt{\frac{(2l_1+1)(2l_2+1)(2l_3+1)}{4\pi}}\times\left[
\begin{matrix}
~l_1~l_2~l_3\\
0~~0~~0
\end{matrix}\right]\left[
\begin{matrix}
l_1~~l_2~~l_3\\
m_1~~m_2~~m_3
\end{matrix}\right]\;.
\end{equation}
The angular quantum numbers $l,l'$ and $L$ rely on specific shells, and the spherical Bessel functions $j_L(qr)$ are also known. $R_{nl}(r)$ denotes the bound electron radial wave-function and $R_{k'l'}(r)$ is the radial wave-function
of the outgoing unbound electrons. The detailed calculation of the radial wave-functions $R_{k'l'}(r)$ and $R_{nl}(r)$ is given in Appendix~\ref{app:ion}.

Now we can numerically determine the value of ionization form factor.
In the fermionic DM absorption by electron target, i.e., $\chi+e\to \nu+e$, the momentum transfer is approximately equal to $q=m_\chi+E_B^{nl}-E_R$ with negative binding energy $E_B^{nl}<0$. We should note that the radial integral in Eq.~(\ref{eq:ionFF}) is complicated in general and the convergence speed depends on different orbitals. It means that we should determine the iteration number of every orbital. In fact, for inner shells 1s, 2s, 2p, $\cdots$, 4s, the electron energies on these orbitals are lower than those of 4p, 4d, 5s and 5p shells. Thus, their radial wave-functions decrease faster (see Fig.~\ref{fig:Rnl} in Appendix~\ref{app:ion}), and the convergence of numerical calculation acts quickly. To efficiently evaluate the integral for the outer shells, we replace Eq.~(\ref{eq:ionFF}) by the following form~\cite{ES2012,1508Lee}
\begin{eqnarray}
\vert f_{\rm{ion}}^{nl}(k',q)\vert^2=\frac{(2l+1){k'}^2}{4\pi^3q}\int_{\vert k'-q\vert}^{\vert k'+q \vert}{kdk\vert\chi_{nl}(k)\vert^2}\;,
\label{equ:A23}
\end{eqnarray}
where $\chi_{nl}(k)$ is an analytical function in momentum space~\cite{2009JK,2021Cao}
\begin{eqnarray}
\begin{split}
\chi_{nl}(k)&=4\pi i^l\int{drr^2R_{nl}(r)j_L(kr)}\\
&=\sum_j C_{nlj}2^{n_{lj}-l}\left(\frac{2\pi a_0}{Z_{lj}}\right)^{3/2}\left(\frac{ika_0}{Z_lj}\right)^l\times\frac{\Gamma(n_{lj}+l+2)}{\Gamma(l+3/2)\sqrt{(2n_{lj})!}}\\
&\times{}_2F_1\left[\frac{1}{2}(n_{lj}+l+2),\frac{1}{2}(n_{lj}+l+3),l+\frac{3}{2},-\left(\frac{a_0k}{Z_{lj}}\right)^2\right]\;.
\end{split}
\label{equ:A24}
\end{eqnarray}
In this way we can easily obtain the ionization form factor for different recoil energies.
However, it does not work well for the case of light DM ($\lesssim 50$ keV) scattering. In our content, as the DM initial momentum is negligible and we assume $E_R-E_B^{nl}\ll m_\chi$, one has $q=m_\chi+E_B^{nl}-E_R\sim m_\chi\gg 0$.
Thus, for larger DM mass and lower recoil energy, the calculation based on the above formula is more precise.

\begin{figure}[H]
\centering
\includegraphics[scale=0.6]{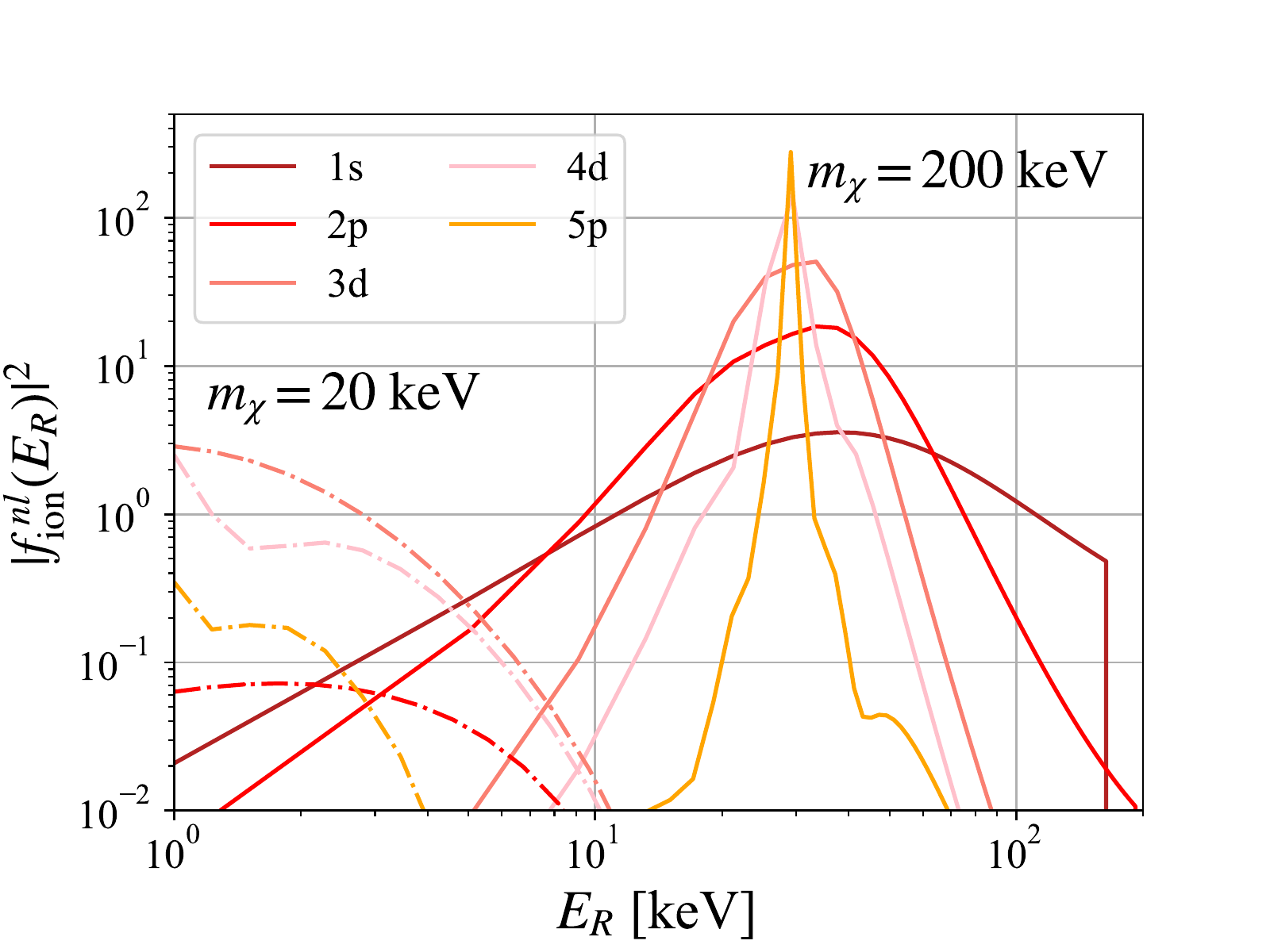}
\caption{The ionization form factors of some electron orbitals with $m_\chi=20$ keV and 200 keV. }
\label{fig:ionFF}
\end{figure}

Thus, to ensure both efficiency and precision, we compute the ionization form factor for different DM masses in different ways. When $m_\chi < 50~\rm{keV}$, we numerically evaluate Eq.~(\ref{eq:ionFF}) by modifying the public code DarkARC from Ref.~\cite{2020RC}. For $m_\chi > 50~\rm{keV}$, we use Eq.~(\ref{equ:A23}) to analytically calculate the form factor. We display the obtained ionization form factor of different orbitals in Fig.~\ref{fig:ionFF}. One can see that the outer orbital shells dominate at $m_\chi$ = 200 keV, while the contributions from shells 1s and 2p are negligible compared to the outer ones. For the case of $m_\chi$ = 20 keV, although the outmost electron does not dominate anymore, there is still a distinct difference between the form factors of the inner and outer orbitals.

In fact, there is also an enhancement due to the attractive potential around the nucleus which results in a larger phase space. This effect is attributed to the difference between the exact and the free (under the plane wave approximation) wave-functions. We compensate it by an extra factor of $E_R$ and $Z_{\rm{eff}}$, usually called Fermi factor $F(E_R,Z_{\rm{eff}})=\vert\Psi_{\rm{exact}}(0)/\Psi_{\rm{free}}(0)\vert^2$~\cite{ES2012}.
It can be given by the following form in non-relativistic limit
\begin{eqnarray}
F(E_R,Z_{\rm{eff}})=\frac{2\pi\xi}{1-e^{-2\pi\xi}}\;,~~~\xi=\alpha Z_{\rm{eff}}\sqrt{\frac{m_e}{2E_R}}\;,
\end{eqnarray}
which tends to unity for large recoil energy.
Actually, this contribution to the ionization form factor is negligible for inner shells.

\begin{figure}
\vspace{0.2cm}
\subfigcapskip=-5pt
\begin{minipage}[t]{0.46\textwidth}
\centering
\includegraphics[scale=0.5]{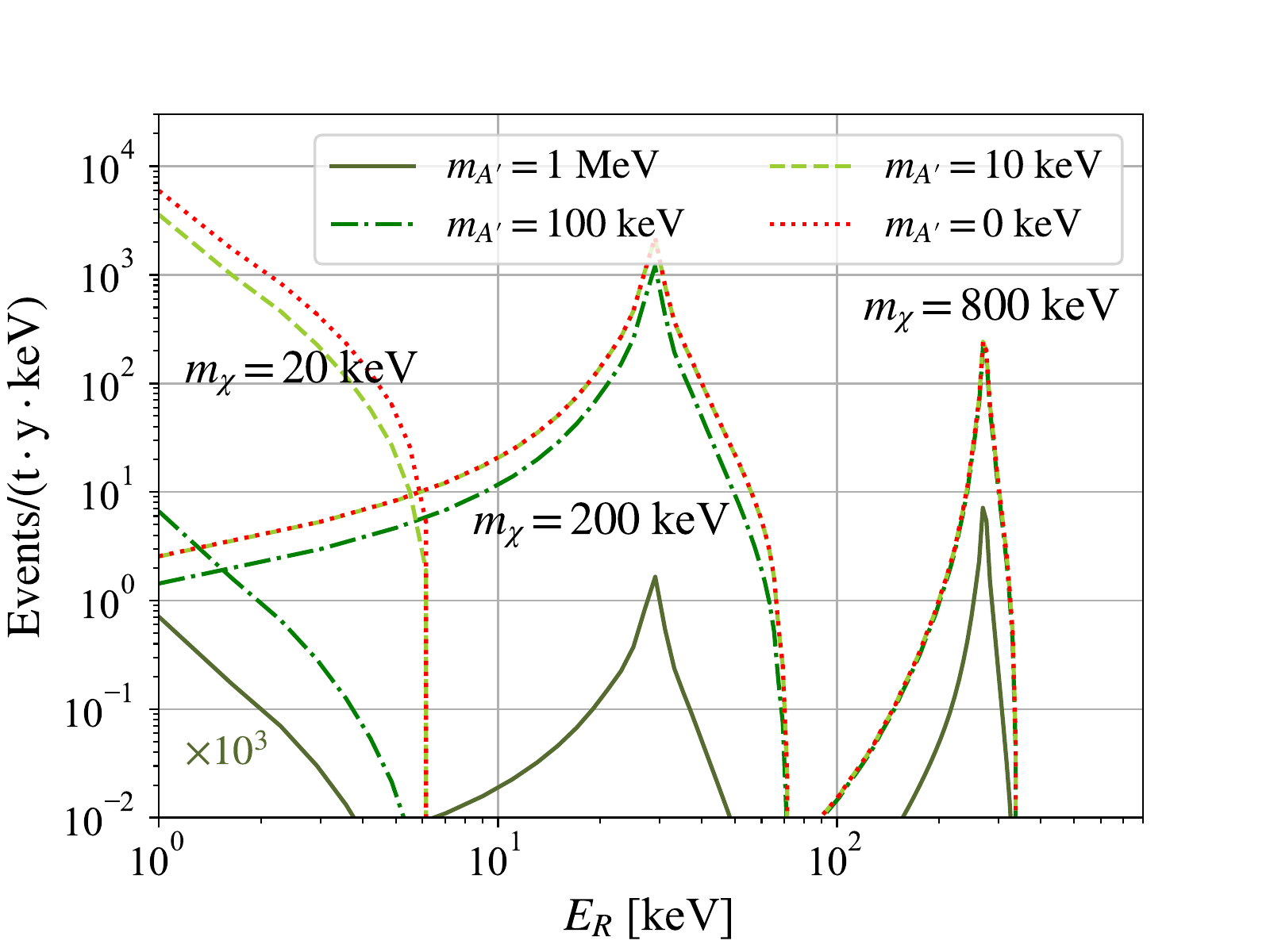}
\end{minipage}
\begin{minipage}[t]{0.65\textwidth}
\centering
\includegraphics[scale=0.5]{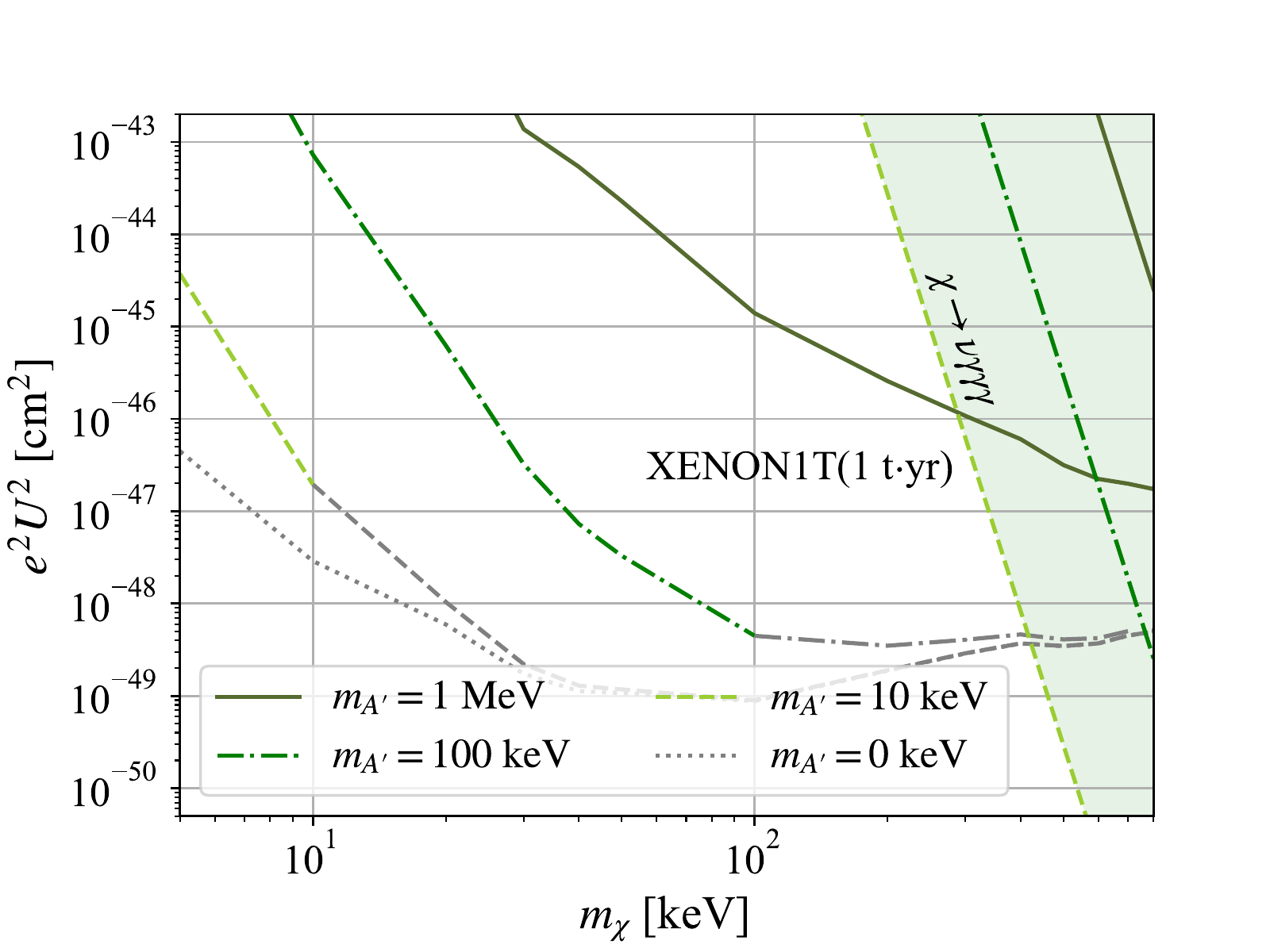}
\end{minipage}
\caption{Left: Ionization rate with different masses of DM and dark photon $m_{A'}$. The coupling $e^2U^2$ is fixed at $10^{-45}~\rm{cm^2}$. The curve corresponding to $m_\chi=20~{\rm keV}$ and $m_{A'}=1~{\rm MeV}$ has been multiplied by a factor of $10^3$. Right: The projected constraint on $e^2U^2$ from the electron ionization of $^{131}{\rm Xe}$ with an exposure of one tonne$\cdot$year. The shaded region has been ruled out by DM lifetime.
}
\label{fig:eCon}
\end{figure}

We show the ionization rate of fermionic DM absorption by electron targets in $^{\rm{131}}\rm{Xe}$ in the left panel of Fig.~\ref{fig:eCon}. The coupling $e^2U^2$ is fixed to be $10^{-45}~\rm{cm^2}$ and we take different masses of DM and dark photon for illustration.
It turns out that lighter DM favors softer spectrum and lighter dark photon enhances the ionization rate.
Note that the step function $\Theta(q)$ in Eq.~(\ref{eq:ediffxsec}) forces the recoil energy range to be $q=m_\chi+E_B^{nl}-E_R>0$ with $E_B^{nl}$ being the binding energy. Thus, the differential scattering rate is truncated at high recoil energies and the cutoff depends on both the DM mass and the binding energy of electron. On the other hand, the profile of ionization rate is determined by the ionization form factor $\vert f_{\rm{ion}}^{nl}(k',q)\vert^2$. We find that each ionization rate with $m_\chi > 50$ keV has a peak of the maximal rate for $E_R > 1$ keV.
Given the ionization form factor, we find that the ionization rate has a broader distribution with respect to the recoil energy than that in DM-nucleon scattering. This signal can thus be searched for through the photoelectron
signature in the detector.

The right panel of Fig.~\ref{fig:eCon} shows the projected constraint on $e^2U^2$ from the electron ionization of $^{131}{\rm Xe}$ with an exposure of one tonne$\cdot$year. We again assume at least 10 events to be observed. The recoil energy has been integrated over from 1 keV to the cutoff. The shaded region has been ruled out by DM lifetime.
One can see that, for extremely small dark photon mass, the limit of the coupling $e^2U^2$ can be pushed down to $10^{-49}$ cm$^2$. Note that the 2-body decay $\chi\to \nu A'$ is allowed if $m_{A'} < m_\chi$. However, this decay has no dependence on the kinetic mixing $\epsilon$ and is parametrically different from the scattering cross section in Eq.~(\ref{eq:ediffxsec}). Despite of this, the inclusion of this additional channel for DM decay would make the constraint more severe for the region of $m_{A'} < m_\chi$. In the right panel of Fig.~\ref{fig:eCon} we use gray color lines to emphasize that the DM mass region of $m_\chi > m_{A'}$ gets more constrained.

\section{Conclusions}
\label{sec:Con}

The fermionic DM absorption by nucleus or electron
targets provides a distinctive signal to search for sub-GeV DM. We consider a Dirac fermion DM charged under a dark gauge group $U(1)'$ and the magnetic dipole operator. The DM field mixes with right-handed neutrino and interacts with the ordinary electromagnetic current via the kinetic mixing term of gauge fields. As a result, the incoming DM is absorbed and converted into neutrino in final state through the dipole-charge interaction.

For the DM absorption by nucleus, the recoil energy spectrum exhibit a peak at $m_\chi^2/2m_N$ for each target isotope. XENON1T can probe the DM mass above 27 MeV and the projected constraint on the inelastic DM-nucleon cross section becomes $10^{-49}$ cm$^2$. CRESSTIII with lower energy threshold would be sensitive to the DM mass around 2 MeV. We also check that the contribution from the nuclear magnetic dipole is negligible for $^{131}{\rm Xe}$ target.

However, the peak-like signature would be difficult to be detected due to the limited energy resolution of detectors.
Moreover, the DM lifetime constraint becomes quite severe for the sensitive region of the inelastic DM-nucleus scattering. These issues motivate us to explore lighter DM through the scattering between DM and bound electrons.

The absorption of DM by bound electron target induces ionization signal and is sensitive to sub-MeV DM mass below 100 keV. The involvement of the ionization form factor spreads out the localized recoil energy. Lighter DM favors softer spectrum and lighter dark photon enhances the ionization rate.
For extremely small dark photon mass, the limit of coupling $e^2 U^2$ can reach as small as $10^{-49}$ cm$^2$.
\\


{\bf{Note Added:}}
During the completion of this work, a study~\cite{EFT} appeared and investigated the fermionic absorption by electron in the framework of effective field theories.

\section*{ACKNOWLEDGMENTS}
T.L. would like to thank Wei Chao for useful discussion.
T.L. is supported by the National Natural Science Foundation of China (Grant No. 11975129, 12035008) and ``the Fundamental Research Funds for the Central Universities'', Nankai University (Grant No. 63196013). J.L. is supported by the National Natural Science Foundation of China (Grant No. 11905299), Guangdong Basic and Applied Basic Research Foundation (Grant No. 2020A1515011479), the Fundamental Research Funds for the Central Universities, and the Sun Yat-Sen University Science Foundation.

\appendix

\section{Nuclear magnetic dipole form factor $F_D(q)$}
\label{app:FD}
The nuclear magnetic dipole form factor arises from the following matrix element which includes the spin-dependent interactions for $\chi-N$ scattering~\cite{Ressell:1997kx}
\begin{eqnarray}
\mathcal{M}=C\left\langle N|a_p\boldsymbol S_p+a_n \boldsymbol S_n|N \right\rangle \cdot \boldsymbol s_\chi\;,
\label{equ:C1}
\end{eqnarray}
where $C$ is the normalization constant, $\boldsymbol S_{n,p}$ are the total nuclear spin operators for neutron $n$ and proton $p$, $\boldsymbol s_\chi$ is the spin operator for DM, and $a_{n,p}$ are the coupling constants for the interactions between DM and nucleon.
In general, the nuclear magnetic moment can be written as~\cite{1992JE}
\begin{eqnarray}
\mu_N = \left\langle N|g_n^s\boldsymbol S_n+g_n^l\boldsymbol L_n+g_p^s \boldsymbol S_p+g_p^l \boldsymbol L_p|N \right\rangle\;,
\label{equ:C2}
\end{eqnarray}
where $\boldsymbol L_{n,p}$ are the orbital angular momentum operators, $g_{n,p}^{l,s}$ are the factors weighting the contributions from different angular momentum operators. For the standard free particles, the $g$ factors are $g_n^s=-3.826$, $g_n^l = 0$, $g_p^s = 5.586$, $g_p^l =1$ (in unit of nuclear magneton $\mu_N$)~\cite{2009PT}. Here we use the effective $g$ factors from the linear least-squares (LLS) fit, i.e.,
$g_n^s=-3.370$, $g_p^s=3.189$, $g_n^l=0.01903$, $g_p^l=1.119$.
For $^{\rm{131}}\rm{Xe}$, after taking the spin and orbital angular momentum matrix elements of proton and neutron from Ref.~\cite{2009PT} and plugging them into Eq.~(\ref{equ:C2}), one can obtain that the ratio of the contributions from orbital and spin angular moments is approximately 0.4 : 0.6 (normalized to 1). According to Ref.~\cite{Chang:2010en}, the nuclear magnetic dipole moment form factor is
\begin{eqnarray}
F_D^2(q)=\left(0.4\frac{L(q)}{L(0)}+0.6\sqrt{\frac{S(q)}{S(0)}}~\right)^2\;,
\end{eqnarray}
where the structure function $L(q)$ and $S(q)$ can be obtained in terms of the spin-independent form factor $F_{\rm{mass}}(q)= L(q)/L(0)$ and the spin-dependent form factor $F_{\rm{spin}}(q)^2=S(q)/S(0)$.

To get a complete analytical form of $F_D(q)$, we need to determine the form factors $F_{\rm{mass}}(q)$ and $F_{\rm{spin}}(q)$. Generally speaking, they are the encapsulation of the matrix element of the nucleon level operators between the nuclear states $|N\rangle$. The form factor $F_{\rm{mass}}(q)$ is defined as  \begin{eqnarray}
F_{\rm{mass}}(q) \equiv \int{d^3\boldsymbol x e^{-i\boldsymbol q\cdot \boldsymbol x}\frac{\rho_n(\boldsymbol x)}{m_n}}\;,
\end{eqnarray}
where $\rho_n(\boldsymbol x)$ is the mass density of nucleons in nucleus.
There are numerous proposed mass density distributions. A commonly used one is proposed by Helm~\cite{Helm:1956zz}
\begin{eqnarray}
\rho_n(r)=\rho_0\left[1+ {\rm exp} {\left(\frac{r-c}{a}\right)} \right]^{-1}\;,
\end{eqnarray}
where $a$ and $c$ are the parameters obtained by using two-parameters least-squares fit.
This distribution has the advantage of yielding an analytical form, that is the so-called Helm form factor
\begin{eqnarray}
F_{\rm mass}(q)\to \frac{3j_1(qr_n)}{qr_n}e^{-(qs)^2/2}\;,
\end{eqnarray}
where $r_n$ has a good fit related to the atom mass number $A$: $r_n\approx 1.14A^{1/3}$, and $s\approx 0.9$ fm.
For the spin-dependent form factor, the spin structure function $S(q)$ is the following linear superposition which includes pure isoscalar $S_{00}$, isovector $S_{11}$ and interference term $S_{01}$
\begin{eqnarray}
S(q) = a_0^2S_{00}(q) + a_1^2S_{11}(q) + a_0a_1S_{01}(q)\;,
\end{eqnarray}
where $a_0~(a_1)$ is the isoscalar (isovector) projection of $a_n$ and $a_p$, and they have a fixed ratio by using EMC (European Muon Collaboration) values for proton: $a_0/a_1=0.297$~\cite{Re1993}. We normalize it by requiring $S(0)=1$, and then one has $a_0 = 2.775$ and $a_1 = 9.341$, or equivalently, $a_p=6.058,~a_n=-3.283~(a_0=a_p+a_n, a_1=a_p-a_n)$. The components $S_{ij}(q)$ of $^{131}\rm{Xe}$ can be found in Ref.~\cite{Ressell:1997kx}. Note that the atomic number of xenon ($Z = 54$) is between those of cesium ($Z = 55$) and iodine ($Z = 53$) which are both proton-odd. For neutron-odd isotope $\rm^{131}Xe$, according to the suggested values in Ref.~\cite{Ressell:1997kx}, it follows that $\vert  \langle \boldsymbol S_p\rangle\vert\ll\vert  \langle \boldsymbol S_n\rangle\vert$.
Thus, the dipole moment in xenon is dominated by neutron spin $\boldsymbol S_n$ unlike the cesium and iodine whose dipole moments are dominated by $\boldsymbol S_p$.

\section{The calculation of the radial wave-functions in the electron ionization form factor $\vert f_{\rm{ion}}^{nl}(k',q)\vert^2$}
\label{app:ion}

Here we give a general method to numerically determine the ionization form factor for different shells and DM masses.
According to the general form of Eq.~(\ref{eq:ionFF}), we have known that $l,l'$ and $L$ are the angular quantum numbers relying on specific shells, and $j_L(qr)$ is also a known spherical Bessel function. Thus, only the radial wave-functions $R_{k'l'}(r)$ and $R_{nl}(r)$ are unknown. In this section, we mainly discuss how to calculate these two functions.

\subsection{The radial wave-function $R_{k'l'}(r)$}

To obtain the radial wave-function $R_{k'l'}(r)$ of outgoing unbound electrons, one needs to solve the radial Schr\"{o}dinger equation with a central potential $Z_{\rm{eff}}(r)/r$. In order to get $Z_{\rm{eff}}(r)/r$, we take the assumption that it can be approximated as the central potential corresponding to the bound state of hydrogen-like atom. Thus, the function $Z_{\rm{eff}}$ becomes a radial independent factor whose values are determined by the following Table~\ref{tab:Zeff} from Ref.~\cite{BUNGE1993113}.

\begin{table}[H]
\resizebox{\linewidth}{!}{\renewcommand\arraystretch{1.5}
\centering
\begin{tabular}{cccccccccccc}
\hline
orbital &1s&2s&2p&3s&3p&3d&4s&4p&4d&5s&5p\\
\hline
$\vert E^{nl}_B\vert$ (keV)  & 33.3174 & 5.1522 & 4.8377 & 1.0932 & 0.9584 & 0.7107 & 0.2138 & 0.1635 & 0.0756 & 0.0257 & 0.0124 \\
$Z_{\rm{eff}}$ & 49.5   & 38.93  & 37.72  & 26.90   & 25.18   & 21.69  & 15.86   & 13.87   & 9.43   & 6.87   & 4.77 \\
\hline
\end{tabular}}
\caption{Binding energy and effective charge of each shell in Xenon.}
\label{tab:Zeff}
\end{table}

\begin{figure}[H]
\centering
\vspace{0.2cm}
\subfigtopskip=2pt
\subfigbottomskip=2pt
\subfigcapskip=-5pt
\subfigure[~Orbital 1s]{
\label{Rkl1s}
\includegraphics[width=0.45\linewidth]{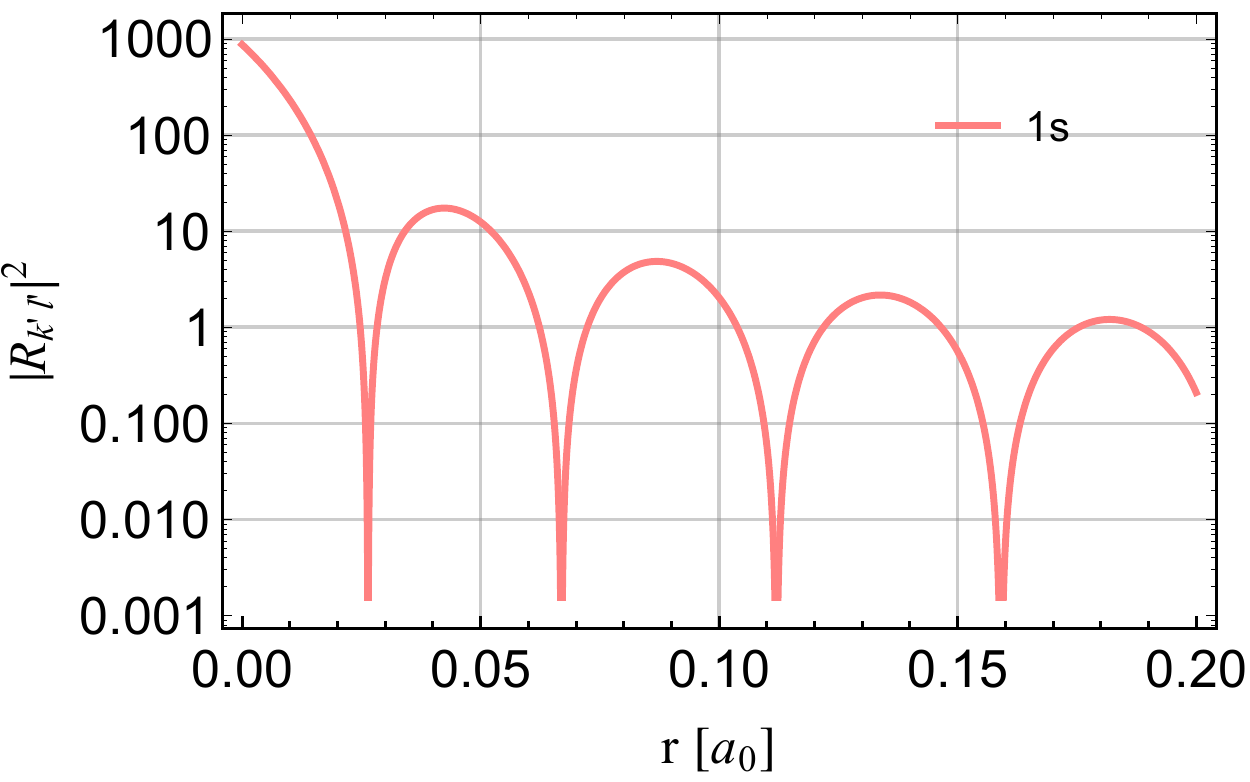}}
\
\subfigure[~Orbital 2p]{
\label{Rkl2p}
\includegraphics[width=0.45\linewidth]{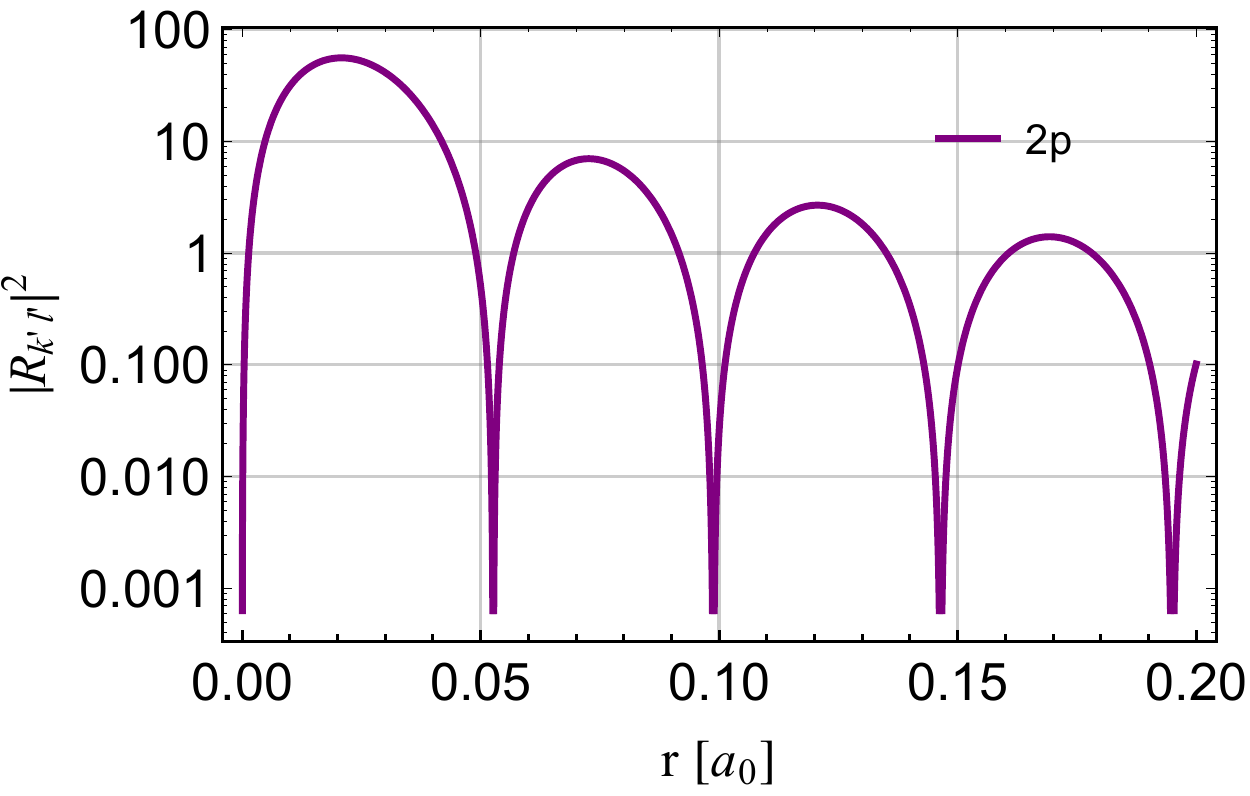}}
\subfigure[~Orbital 3d]{
\label{Rkl3d}
\includegraphics[width=0.45\linewidth]{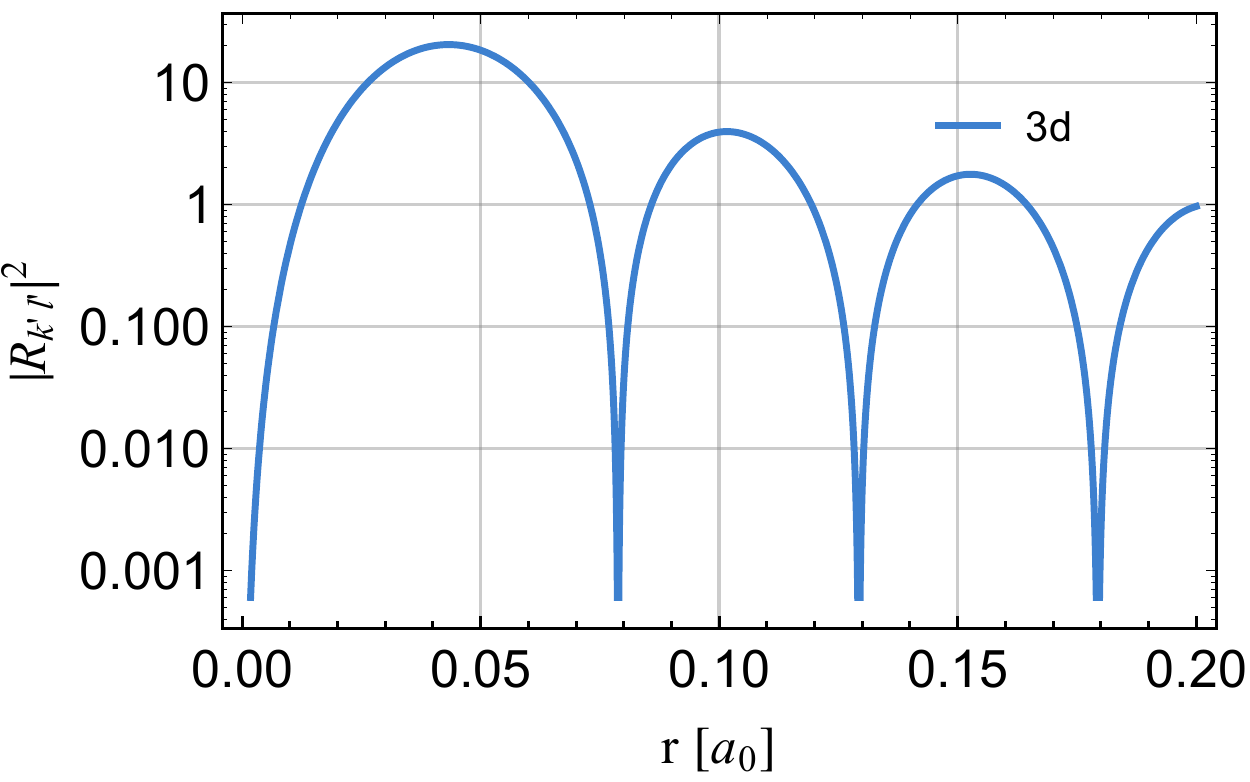}}
\
\subfigure[~Orbital 4p]{
\label{Rkl4p}
\includegraphics[width=0.45\linewidth]{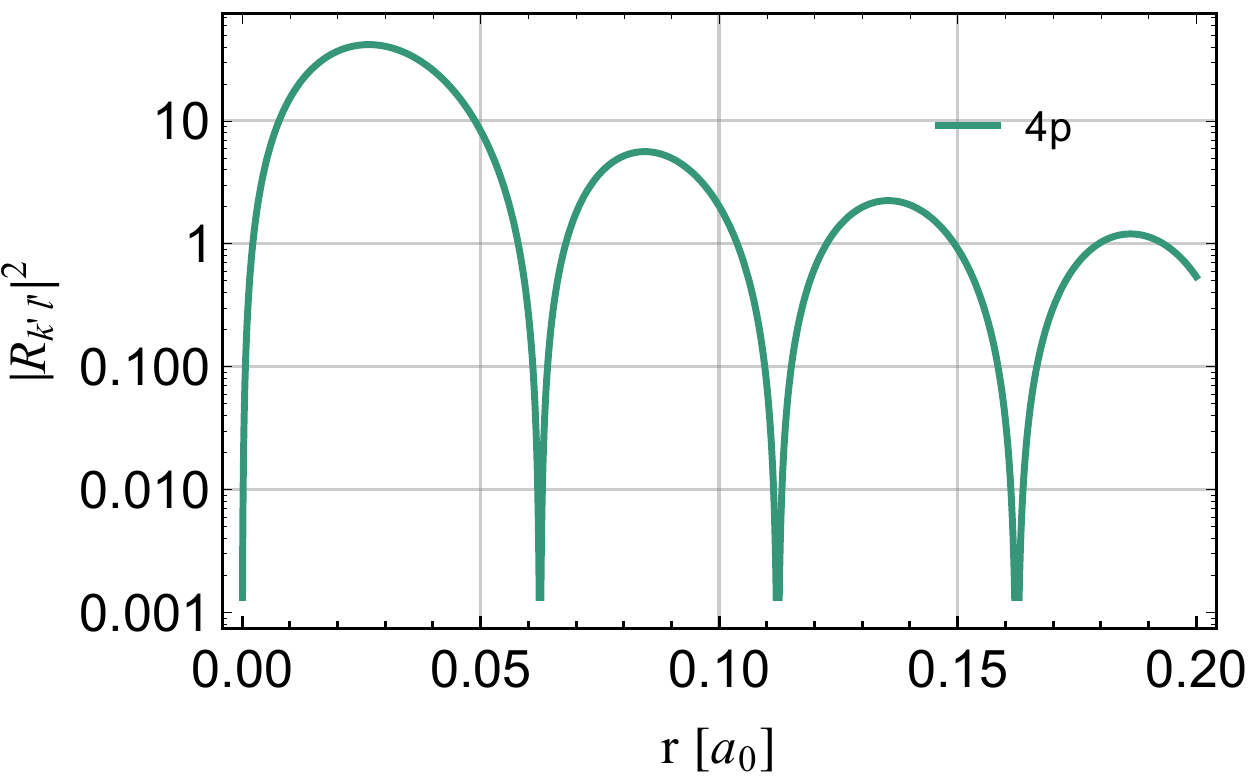}}
\caption{The radial wave-function $R_{k'l'}(r)$ of the outgoing electron with recoil energy $E_R=50$ keV for illustration.}
\label{fig:Rkl}
\end{figure}

$Z_{\rm{eff}}$, the effective charge felt by the scattered electron, is then determined by the binding energy of different orbitals $\displaystyle Z_{\rm{eff}}^{nl}=\sqrt{\vert E_B^{nl}\vert/13.6~\rm{eV}}\times n$ with $E_B^{nl}$ being the binding energy. The general solution with positive energy continuum for radial Sch\"{o}dinger equation of hydrogen-like atom is given by~\cite{2018DarkSide,Bethe:1957ncq}
\begin{eqnarray}
R_{k'l'}(r)&=&\frac{(2\pi)^{3/2}}{\sqrt{V}}(2k'r)^{l'} \frac{\sqrt{\frac{2}{\pi}}\left\vert\Gamma(l'+1-\frac{iZ_{\rm{eff}}}{k'a_0})\right\vert e^{\pi Z_{\rm{eff}}\over{2k'a_0}}}{(2l'+1)!}\nonumber \\
&&\times e^{-ik'r}{}_1F_1(l'+1+\frac{iZ_{\rm{eff}}}{k'a_0},2l'+2,2ik'r)\;,
\end{eqnarray}
where ${}_1F_1(a,b,z)$ is the first kind confluent hypergeometric function, and $a_0$ denotes the Bohr radius~\footnote{The Bohr radius in natural unit is $a_0=4\pi\epsilon_0 h^2/m_ee^2=r_e\alpha^{-2}=0.529177210903(80)\times10^{-10}$ m ($m_{\rm{nucleus}}=\infty$)~\cite{Zyla:2020zbs}.}.
It should be noticed that the extra volume factor $1/\sqrt{V}$ can be eventually cancelled by the integral of space. It means that once the momentum of the outgoing electron $k'$ and the orbital quantum number $l'$ are determined, we can evaluate the wave-function $R_{k'l'}$ at any radii. At the same time, due to the asymptotic approximation, the wave-function value is more accurate at large radii.
In Fig.~\ref{fig:Rkl} we show the numerical results of radial wave-function $R_{k'l'}(r)$ corresponding to 1s, 2p, 3d, 4p shells, respectively.

\subsection{The radial wave-function $R_{nl}(r)$}

\begin{figure}[H]
\centering
\vspace{0.2cm}
\subfigtopskip=2pt
\subfigbottomskip=2pt
\subfigcapskip=-5pt
\subfigure[~Orbital 1s]{
\label{Rkl1s}
\includegraphics[width=0.45\linewidth]{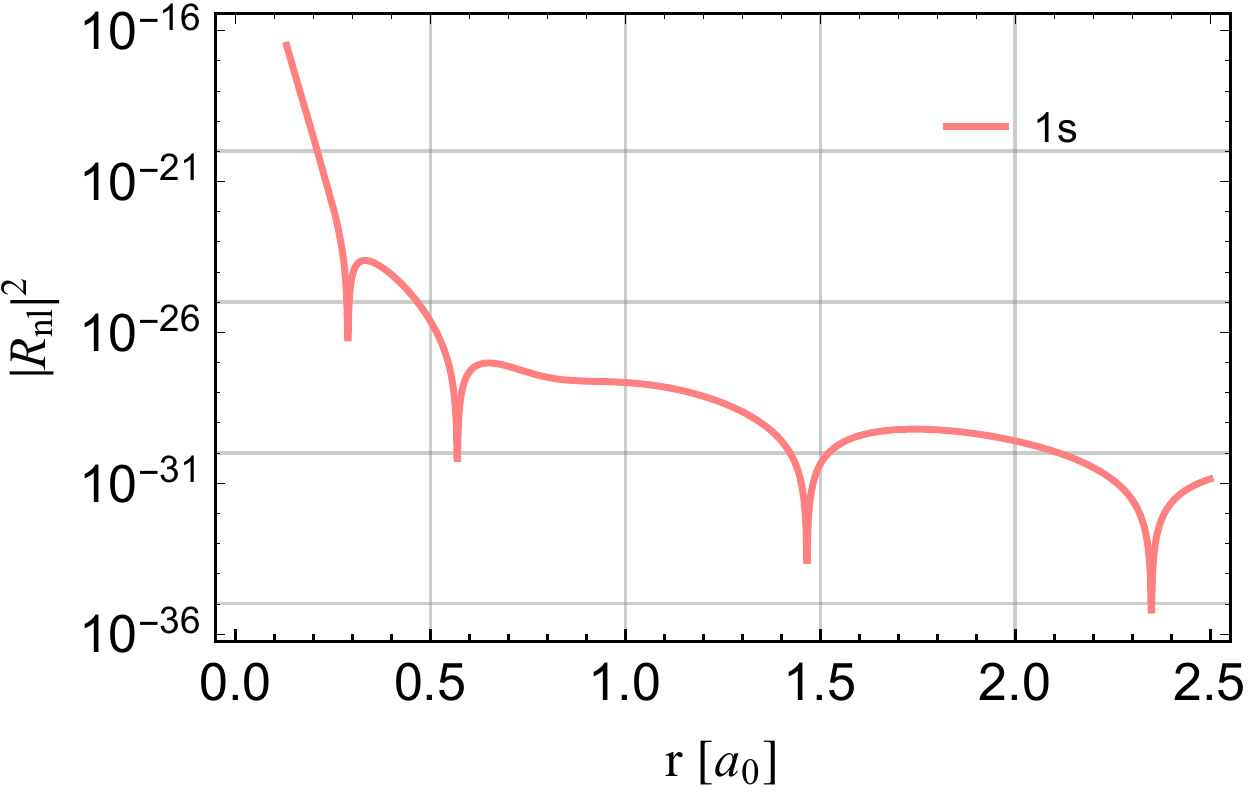}}
\
\subfigure[~Orbital 2p]{
\label{Rkl2p}
\includegraphics[width=0.45\linewidth]{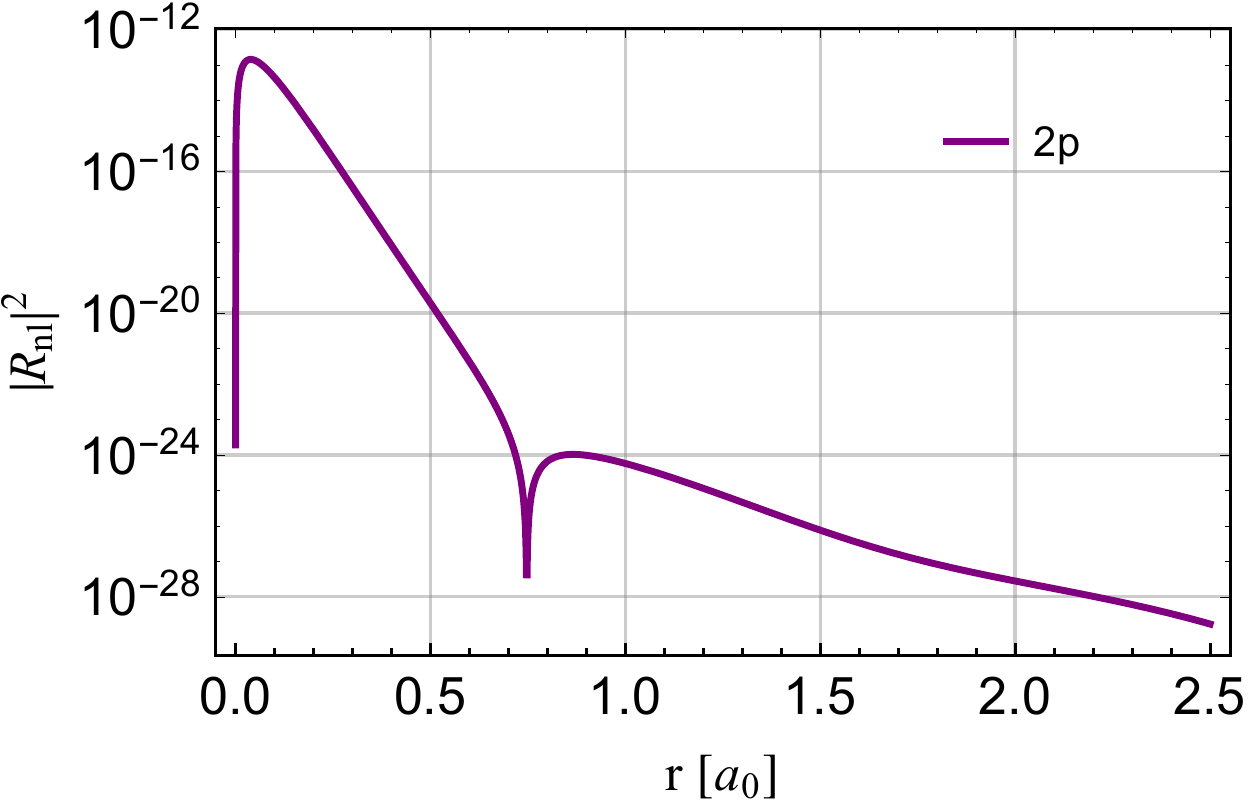}}
\subfigure[~Orbital 3d]{
\label{Rkl3d}
\includegraphics[width=0.45\linewidth]{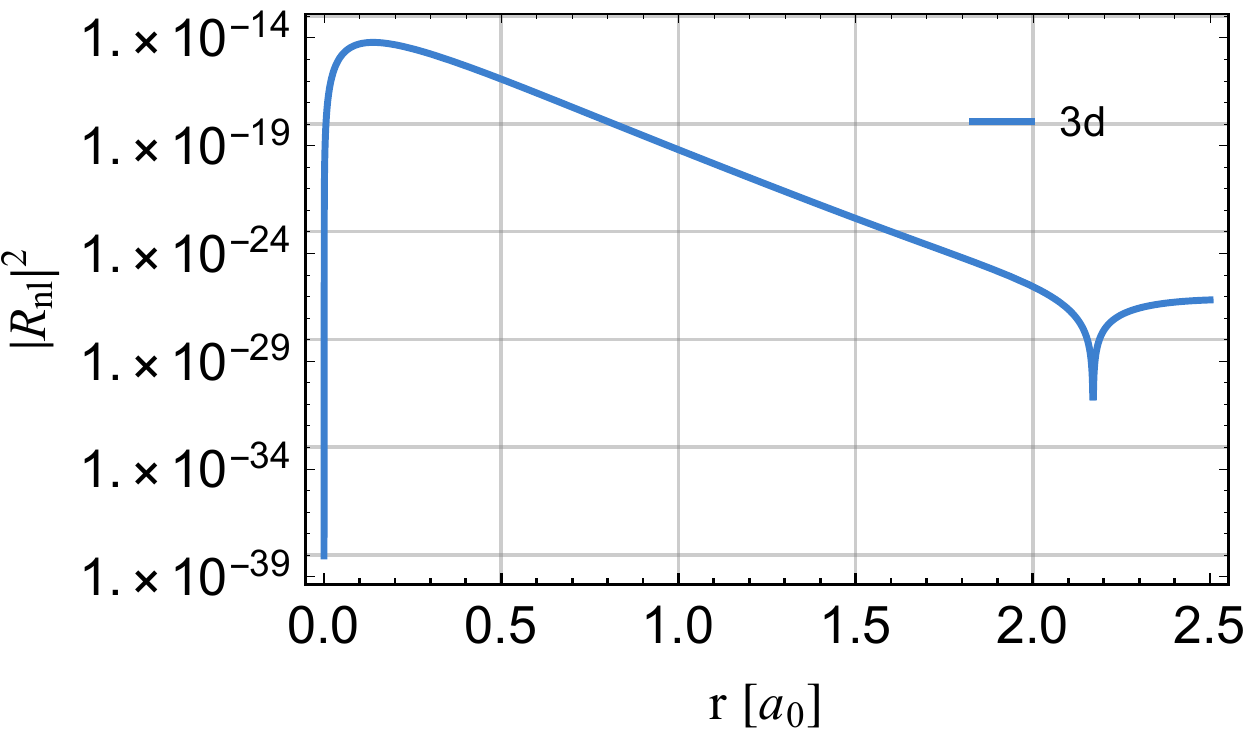}}
\
\subfigure[~Orbital 4p]{
\label{Rkl4p}
\includegraphics[width=0.45\linewidth]{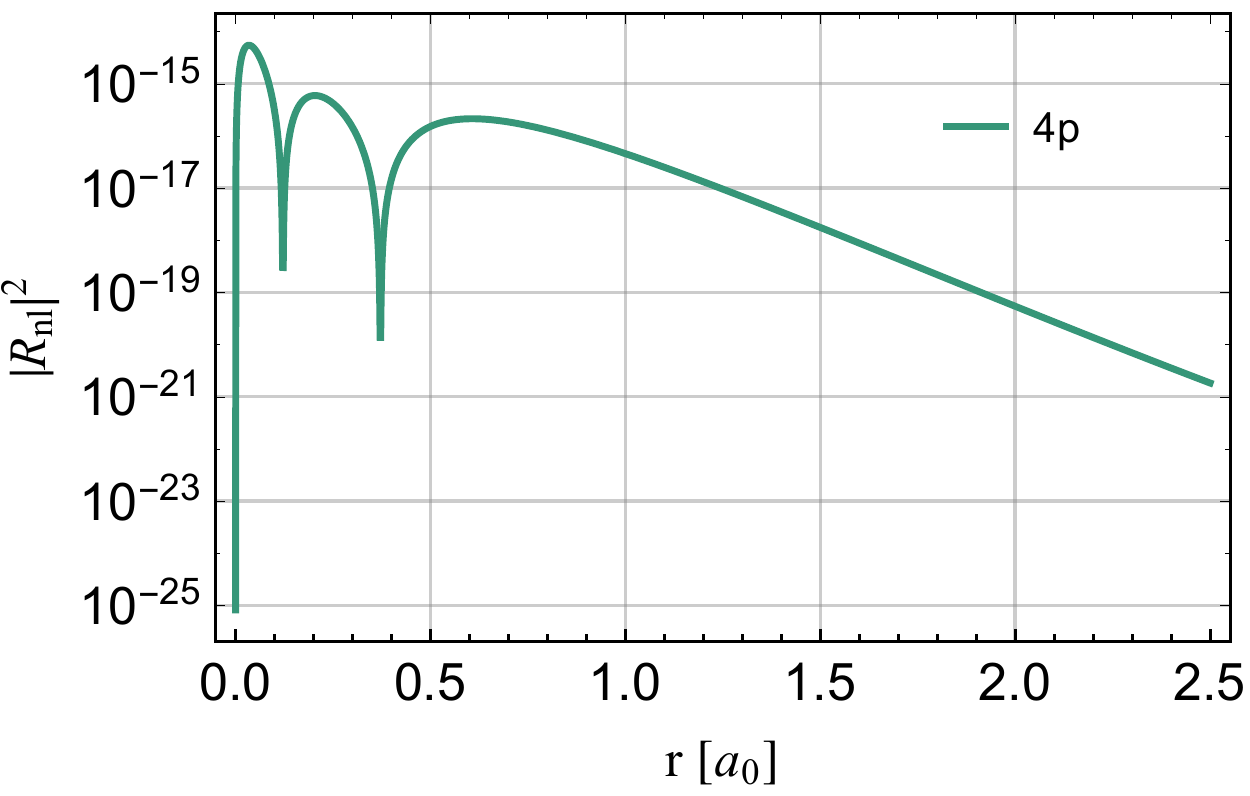}}
\caption{The radial wave-function $R_{nl}(r)$ of the bound electron outside the xenon nucleus.}
\label{fig:Rnl}
\end{figure}

The orbital wave-function $R_{nl}$ on the $(n,l)$ shell can be given as a linear combination of the Slater-type orbitals via the Roothaan-Hartree-Fock (RHF) approximation~\cite{BUNGE1993113}. In the RHF method, the atomic orbitals of xenon can be obtained as a finite superposition of analytical radial function
\begin{equation}
R_{nl}(r)=a_0^{-3/2}\sum_jC_{jln}
\frac{(2Z_{jl})^{n'_{jl}+1/2}}{\sqrt{(2n'_{jl})!}}\times\left(\frac{r}{a_0}\right)^{n'_{jl}-1}{\rm exp}\left(-Z_{jl}\frac{r}{a_0}\right)\;,
\end{equation}
where the coefficients $C_{jln}$, $Z_{jl}$ and $n'_{jl}$ are all obtained from Ref.~\cite{BUNGE1993113}. Similarly, in Fig.~\ref{fig:Rnl}, we show the results of radial wave-function $R_{nl}(r)$ for 1s, 2p, 3d, 4p shells, respectively.

\bibliographystyle{JHEP}
\bibliography{refs}

\end{document}